\documentclass[preprintnumbers,twocolumn,article,amsmath,amssymb,floatfix,9pt,prd,superscriptaddress,nofootinbib]{revtex4-1}
\usepackage[utf8,latin1]{inputenc}
\usepackage{graphicx}
\usepackage{dcolumn}
\usepackage[dvipsnames]{xcolor}
\usepackage[T1]{fontenc}

\usepackage{mathrsfs}  
\usepackage{cases}
\usepackage{bm}
\usepackage{academicons}
\usepackage{mathtools, nccmath}
\usepackage{fancyhdr}
\usepackage{tikz,xcolor}

\usepackage{tensor}
\usepackage[normalem]{ulem}
\usepackage{lipsum}
\usepackage{soul}
\usepackage{cancel}
\usepackage{stackengine,scalerel}
\usepackage{mathabx}
\usepackage{hyperref}
\usepackage{tabularx}
\hypersetup{colorlinks, linkcolor={red},citecolor={blue},urlcolor={blue}}  

\AtBeginDocument{\fontsize{8}{11}\selectfont} 

\renewcommand{\arraystretch}{1.2}

\definecolor{lime}{HTML}{A6CE39}
\DeclareRobustCommand{\orcidicon}{
	\begin{tikzpicture}
	\draw[lime, fill=lime] (0,0) 
	circle [radius=0.16] 
	node[white] {{\fontfamily{qag}\selectfont \tiny ID}};
	\draw[white, fill=white] (-0.0625,0.095) 
	circle [radius=0.007];
	\end{tikzpicture}
	\hspace{-2mm}
}

\fancypagestyle{plain}{%
  \fancyhf{}
  \fancyfoot[C]{\iffloatpage{}{\thepage}}
  }
\foreach \x in {A, ..., Z}{%
	\expandafter\xdef\csname orcid\x\endcsname{\noexpand\href{https://orcid.org/\csname orcidauthor\x\endcsname}{\noexpand\orcidicon}}
}

\begin{document}

\title[Kazakov-Solodukhin Quantum-Deformed Black Holes and Their Observational Signatures: Testing the Deformation Parameter through Black Hole Shadow Measurements of M87$^*$ and Sgr A$^*$]{Observational Constraints on Kazakov-Solodukhin Quantum-Deformed Black Holes from M87$^*$ and Sgr A$^*$ Shadows}

\author{A. Errehymy\orcidA{}}%
\email{abdelghani.errehymy@gmail.com (Corresponding author)}
\affiliation{Astrophysics Research Centre, School of Mathematics, Statistics and Computer Science, University of KwaZulu-Natal, Private Bag X54001, Durban 4000, South Africa}
\affiliation{Center for Theoretical Physics, Khazar University, 41 Mehseti Str., Baku, AZ1096, Azerbaijan}

\author{Y. Khedif\orcidB{}}%
\email{youssef.khedif@gmail.com}
\affiliation{Department of Physics, Faculty of Sciences A\"{i}n Chock, Laboratory of Mechanics and High Energy Physics, University Hassan II, P.O. Box 5366, 20100, Maarif Casablanca, Morocco}

\author{M. Daoud\orcidC{}}%
\email{m$_{}$daoud@hotmail.com}
\affiliation{Department of Physics, Faculty of Sciences, Ibn Tofail University, P.O. Box 133, Kenitra 14000, Morocco}
\affiliation{Abdus Salam International Centre for Theoretical Physics, Miramare, Trieste 34151, Italy}

\author{B. Turimov\orcidD{}}%
\email[]{bturimov@astrin.uz}
\affiliation{Central Asian University, Milliy bog Str. 264, Tashkent, 111221, Uzbekistan}
\affiliation{University of Tashkent for Applied Sciences, Gavhar Str. 1, Tashkent, 100149, Uzbekiston}

\author{S. Usanov\orcidE{}}
\email[]{sm.usanov@kiut.uz}
\affiliation{Kimyo International University in Tashkent, Shota Rustaveli Str. 156, Tashkent 100121, Uzbekistan}

 \author{Z. Yasakov\orcidF{}}
 \email[]{zikrillo87@mail.ru}
\affiliation{Samarkand State University of Architecture and Construction, Lolazor Street 70, 140147, Samarkand, Uzbekistan}

\author{Z. Avezmuratova\orcidI{}}
\email[]{zeboavezmuratova1981@mail.ru}
\affiliation{Department of Physics and Astronomy, Urgench State Pedagogical Institute, Gurlan Str.1-A, Urgench 220100, Uzbekistan}

\date{\today} 

\begin{abstract}
{\footnotesize { We explore the Kazakov-Solodukhin quantum-deformed black hole spacetime, characterized by a single deformation parameter \( \eta \) that encodes quantum corrections to the classical Schwarzschild solution. The model preserves the correct general-relativistic limit as \( \eta \to 0 \), while introducing significant and physically meaningful deviations in the strong-field regime. A central and remarkable feature of the geometry is the regularization of the classical singularity: curvature invariants remain finite near the minimal radius \( r = \eta \), effectively replacing the divergent core with a smooth and well-behaved region. This behavior naturally introduces a minimal length scale into the spacetime structure, offering a geometrically motivated resolution of the singularity problem. The deformation modifies the horizon structure, shifts the event horizon location, and alters the mass-radius relation. It also reduces the surface gravity, leading to a lower Hawking temperature and a slower evaporation process, thereby enhancing the thermodynamic stability of the black hole. Photon dynamics are correspondingly affected, resulting in a displaced photon sphere and modified strong-lensing characteristics. While the shadow remains perfectly circular due to spherical symmetry, its size depends sensitively on \( \eta \). Observational constraints can be expressed through
\(
\left| R_{sh}(\eta) - R_{obs} \right| \leq \Delta R_{obs},
\)
which places an upper bound on the deformation parameter. In the weak-field limit, the deflection angle acquires a quadratic correction proportional to \( \eta^2 \), ensuring consistency with precision tests while allowing potentially detectable deviations in strong-gravity observations. These features make the model both theoretically appealing and observationally testable.
}}
\end{abstract}

\maketitle
\section{Introduction:} \label{Sec:1}
Black holes stand among the most fascinating objects in modern astronomy, combining extreme gravity with rich and complex physical environments. For many years they were considered purely theoretical predictions, but recent observations have transformed them into directly imaged astrophysical sources. A major milestone was achieved by the Event Horizon Telescope (EHT), which delivered the first image of the supermassive black hole in the galaxy Messier (M) 87$^*$ \cite{EventHorizonTelescope:2019dse, EventHorizonTelescope:2019uob, EventHorizonTelescope:2019jan, EventHorizonTelescope:2019ths, EventHorizonTelescope:2019pgp, EventHorizonTelescope:2019ggy}. This groundbreaking result not only revealed the characteristic shadow structure but also provided valuable information about the surrounding emission and magnetic activity. Further studies by the same collaboration examined the magnetic properties of the accretion flow around M87$^*$, showing that the observed structure agrees well with theoretical expectations from general relativistic magnetohydrodynamic models \cite{EventHorizonTelescope:2021srq}. These findings strengthened the link between high-resolution observations and advanced numerical simulations describing matter under extreme gravitational and magnetic conditions. Another remarkable achievement came with the first horizon-scale image of the supermassive black hole at the center of our own galaxy, Sagittarius (Sgr) A$^*$ \cite{EventHorizonTelescope:2022wkp, EventHorizonTelescope:2022apq, EventHorizonTelescope:2022wok, EventHorizonTelescope:2022exc, EventHorizonTelescope:2022urf, EventHorizonTelescope:2022xqj}. Detailed analyses confirmed that the image is dominated by a bright, thick ring-like feature with an angular diameter of about \(51.8 \pm 2.3\) microarcseconds. This result provided strong observational support for long-standing theoretical predictions regarding the appearance of black hole shadows. Taken together, these discoveries highlight an important point: the observable properties of a black hole shadow are not determined solely by spacetime geometry, but can also be influenced by the surrounding accretion disk and its physical characteristics. Therefore, a careful interpretation of shadow measurements must account for the environment around the black hole, as it may play a significant role in shaping the final observed image. Recent years have brought a number of interesting developments in physics and space science. These include studies of multi-asteroid exploration missions using electric sails \cite{Huo:2023}, the first-photon visualization of quantum erasure with hybrid entanglement \cite{Yu:2025}, and new investigations of supermassive black-hole merger rates through gravitational-wave observations \cite{Fang:2025}. Progress has also been achieved in high-temperature quantum anomalous Hall systems \cite{Qiao:2016}, the identification of gamma-ray blazar candidates from multiwavelength observations \cite{Xiang:2026}, and the development of space-time wavefront synchronized terahertz metasurfaces \cite{Zhang:2026}.

Recent years have also seen considerable progress in black hole astrophysics, particularly in exploring the effects of dark matter, cosmic voids, and modified gravity on the properties of compact objects. A number of studies have examined black hole shadows, null geodesics, and related observational signatures in spacetimes influenced by dark matter halos, quintessential fields, and large-scale cosmic structures, providing valuable constraints from Event Horizon Telescope observations of M87$^*$ and Sgr A$^*$ \cite{Errehymy:2025ada, Errehymy:2025pfr, Errehymy:2025nuv, Errehymy:2023xpc}. Other investigations have focused on regular and non-singular black holes, analyzing their orbital dynamics, observational characteristics, and quantum corrections within extended theories of gravity \cite{Errehymy:2025zuj,Errehymy:2025djk,Errehymy:2026ftm}. More recently, high-frequency quasi-periodic oscillations observed in X-ray binaries have been used to place constraints on ultralight bosonic dark matter surrounding black holes \cite{Errehymy:2026hbj}. These developments illustrate the increasing role of astronomical observations in testing theoretical models of gravity and black hole physics. Understanding spacetime singularities has long been one of the central challenges in general relativity (GR). From the earliest developments of the theory, energy conditions were introduced as important tools for analyzing gravitational behavior, and singularity theorems became a cornerstone in testing the internal consistency of GR. However, the prediction of singularities also reveals a deeper limitation: the breakdown of geodesic completeness. When spacetime paths cannot be extended in a well-defined way, it signals that the classical description is no longer sufficient, suggesting that quantum gravitational effects may be required to resolve the issue \cite{Hawking:1970zqf}.

Over the years, many researchers have explored how quantum phenomena behave near event horizons, during gravitational collapse, and in regions close to singular points \cite{Duff:1974ud, Frolov:1981mz}. These investigations indicate that quantum contributions could significantly modify the classical picture and, in some scenarios, may even prevent the formation of true physical singularities at the core of compact objects. Motivated by this possibility, extensive work has examined quantum effects in black hole spacetimes under different theoretical frameworks \cite{Stelle:1976gc, Stelle:1977ry, Biswas:2011ar, Biswas:2013cha, Biswas:2016egy}. In a notable contribution, Hooft \cite{tHooft:1987vrq, tHooft:1988oyr} studied quantum gravity using a semiclassical approach and showed that black hole formation, along with coherent graviton emission, can occur at energies exceeding the Planck scale . Despite such progress, a complete and universally accepted quantum description of black holes is still lacking.  The role of quantum effects in compact objects across various modified gravity frameworks has also been widely discussed in the literature \cite{Battista:2021rlh, Battista:2023glw, Battista:2023iyu, Errehymy:2024mlf, Battista:2024gud, Errehymy:2025llh, Wang:2025fmz, Errehymy:2025djk}.

An interesting direction in this context concerns quantum-deformed solutions of the Schwarzschild metric. Kazakov and collaborators \cite{Kazakov:1993ha} proposed a modified black hole geometry arising from matter-field quantum corrections, arguing that such a deformation could effectively encode quantum fluctuations in the surrounding field. Building on this idea, Konoplya \cite{Konoplya:2019xmn} analyzed the corresponding spacetime using quasinormal mode calculations and demonstrated that introducing quantum deformation leads to a reduction in the black hole shadow radius. Further studies by Xu and co-workers \cite{Lu:2021htd} examined both weak and strong gravitational lensing in the Kazakov-Solodukhin black hole background, revealing potential observational signatures. Because of these intriguing features, the Kazakov-Solodukhin black hole has been widely applied in various research directions \cite{Bezerra:2019qkx, Berry:2021hos, Gao:2021ybp, Javed:2021ymu}. Despite these developments, an important question remains open: can quantum corrections --- often considered primarily at the theoretical level --- produce measurable consequences in astrophysical observations? In this direction, Peng and colleagues \cite{Peng:2020wun} investigated the shadow of a quantum-corrected Schwarzschild black hole surrounded by an optically and geometrically thin accretion disk. Their analysis showed that quantum modifications can alter the structure of the photon sphere and critical curve, leading to a violation of the standard inequality satisfied by asymptotically flat black holes obeying the zero energy condition in Einstein gravity. As a result, the observable appearance of the accretion disk and the associated shadow can differ noticeably from the classical prediction \cite{Peng:2020wun}.

Motivated by these developments, in this work,  we adopt the Kazakov-Solodukhin quantum-deformed black hole \cite{Kazakov:1993ha} as a simple and transparent model to explore possible quantum effects in gravity. The model is attractive because it is controlled by a single parameter \( \eta \), which smoothly connects the Schwarzschild spacetime to a regular, quantum-inspired geometry without changing the asymptotic structure. In this way, the weak-field behavior remains consistent with well-established tests of GR, while meaningful deviations appear only in the strong-gravity region near the horizon. This balance makes the model both theoretically appealing and observationally relevant. The deformation influences the horizon radius, thermodynamic quantities, photon motion, and shadow size, providing clear physical signatures that can be explored using black hole imaging, gravitational lensing, and other high-precision observations. At the same time, current experimental data impose natural limits on the allowed range of \( \eta \), ensuring compatibility with existing measurements. Motivated by these considerations, we systematically analyze the geometric, thermodynamic, and observational properties of the spacetime and examine how the deformation parameter affects measurable quantities in the strong-field regime.

The paper is organized as follows. In Sect. \ref{Sec:2}, we introduce the quantum-deformed spacetime and discuss its geometric structure and main properties. Sect. \ref{Sec:3} is devoted to the thermodynamic aspects of the black hole, where we analyze the Hawking temperature and the associated energy emission. In Sect. \ref{Sec:4}, we study the motion of photons in this background and examine the structure of null trajectories. The weak-field behavior is explored in Sect. \ref{Sec:5}, where we derive the light-bending angle and discuss how it depends on the deformation parameter. The dynamical stability of circular photon orbits is investigated in Sect. \ref{Sec:6}, focusing on the role of the Lyapunov exponent. In Sect. \ref{Sec:7}, we compare the theoretical predictions with available astrophysical observations and discuss the resulting constraints on the model. Finally, Sect. \ref{Sec:8} presents a summary of the main findings and outlines possible future directions.

\section{Kazakov-Solodukhin metric}\label{Sec:2}

Quantum effects can modify the classical Schwarzschild geometry when fluctuations of the spacetime metric are taken into account. In particular, Dmitri Kazakov and Sergey Solodukhin showed in their work \cite{Kazakov:1993ha} that spherically symmetric quantum fluctuations lead to a deformation of the Schwarzschild solution. Within this approach, the four-dimensional Einstein theory effectively reduces to a two-dimensional dilaton gravity model, allowing quantum corrections to be incorporated into the gravitational potential. The resulting spacetime is described by the Kazakov-Solodukhin metric \cite{Kazakov:1993ha}, which takes the form
\begin{equation}\label{Metr}
ds^2 = - f(r)\, dt^2 + f^{-1}(r)\, dr^2 + r^2 \left(d\theta^2 + \sin^2\theta\, d\varphi^2\right).
\end{equation}

In this framework, the effective renormalizable potential \(V(\epsilon)\) is given by
\begin{equation}
V(\epsilon)=e^{-\epsilon}\left(e^{-2\epsilon}-\frac{4\tilde{G}}{\pi}\right)^{-1/2},\label{4}
\end{equation}
where \(\tilde{G} = G_N \ln(\mu/\mu_{*})\). Here \(G_N\) denotes Newton's gravitational constant, while \(\mu\) represents a renormalization scale.

Using this potential, the metric function of the quantum-corrected Schwarzschild geometry can be written as
\begin{equation}
f(r)=-\frac{2M}{r}+\frac{1}{r}\int^r V(\epsilon)\, d\epsilon,\label{2}
\end{equation}
which leads to the explicit expression
\begin{equation}
f(r)=\left(\frac{\sqrt{r^2-\eta^2}}{r}-\frac{2M}{r}\right).\label{lapsefunction}
\end{equation}

The quantity \(\eta^2 = 4\tilde{G}/\pi\) acts as a deformation parameter encoding the strength of the quantum corrections. In the classical limit corresponding to empty space, the potential reduces to \(V(\epsilon)=1\), and the standard Schwarzschild solution is recovered.

\begin{figure}
\centering
\includegraphics[width=7.9cm,height=5.9cm]{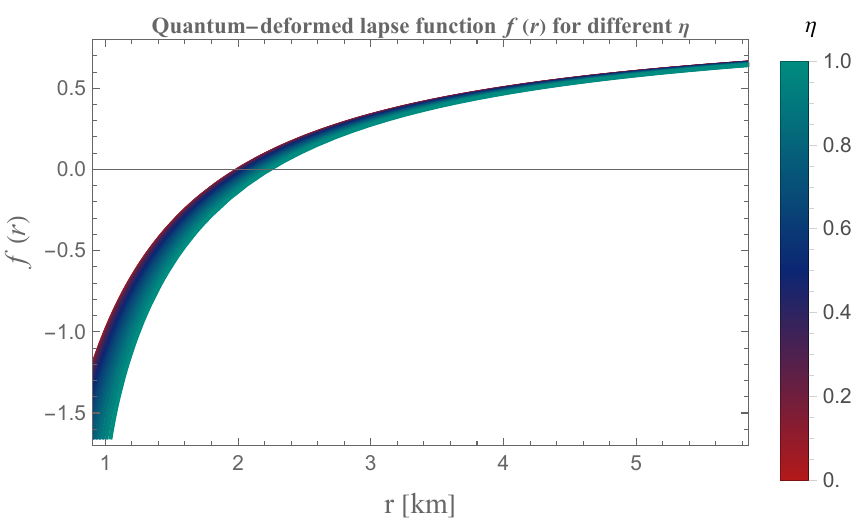}%
    \caption{\scriptsize The figure shows the radial profile of the quantum-deformed lapse function (\ref{lapsefunction}) for \(0 \le \eta \le 1\) with fixed mass \(M=1\). Each curve corresponds to a different value of \(\eta\), as indicated by the color scale. The radial coordinate is taken slightly above \(r=\eta\) to ensure the function remains real. Increasing \(\eta\) modifies both the shape and magnitude of \(f(r)\) relative to the undeformed case (\(\eta=0\)), demonstrating how the deformation parameter influences the spacetime geometry through the lapse function. 
    }    \label{Fig1}
\end{figure}

\begin{figure}
\begin{center}
\includegraphics[width=7.9cm,height=5.9cm]{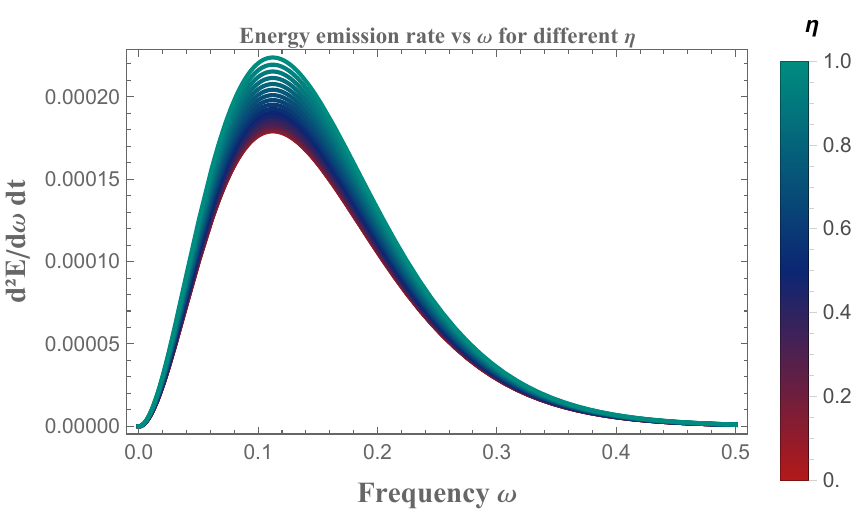}
\end{center}
\caption{\scriptsize The figure presents the energy emission rate \(\frac{d^2E}{d\omega dt} \) as a function of the frequency \( \omega \) for a quantum-deformed black hole with mass \( M=1 \). The geometry is defined by the lapse function (\ref{lapsefunction}). Each curve corresponds to a different value of the deformation parameter \( \eta \in [0,1] \), shown using the same color scheme for consistency. Varying \( \eta \) modifies the horizon structure and thermodynamic properties, leading to noticeable changes in both the amplitude and profile of the emission spectrum.
 }\label{Fig2}
\end{figure}

The Fig. \ref{Fig1} shows how the lapse function \(f(r)\) behaves when the deformation parameter \( \eta \) changes, while the mass is kept fixed at \(M=1\). Each colored curve corresponds to a different value of \( \eta \), as indicated by the color bar. By following the curves from left to right, one can clearly see how the geometry of the spacetime is modified as the deformation parameter grows. A first point that stands out in the plot is that the curves do not start at the same radial position. Instead, each one begins slightly to the right of the previous one. This occurs because the metric is only defined for radii larger than \( \eta \). As the value of \( \eta \) increases, the allowed region of the spacetime shifts outward. In physical terms, this means that the deformation introduces a characteristic length scale that prevents the geometry from extending to arbitrarily small radii. Another important feature can be seen where the curves cross the horizontal axis, which corresponds to the condition \(f(r)=0\). This point marks the location of the event horizon. From the plot it becomes clear that the intersection with the axis moves gradually toward larger values of \(r\) as \( \eta \) increases. In other words, the presence of the deformation parameter effectively pushes the horizon outward, slightly enlarging the region from which light cannot escape. Far from the central object the situation becomes much simpler. At large radial distances the curves tend to merge and follow nearly the same trajectory. This behavior indicates that the influence of the deformation becomes negligible in the weak-gravity region. Consequently, the spacetime retains the usual Schwarzschild behavior at large \(r\), while the modifications introduced by \( \eta \) remain confined mainly to the strong-gravity region close to the horizon. Taken together, the plot suggests that the parameter \( \eta \) primarily affects the inner structure of the spacetime and the position of the horizon, while leaving the asymptotic gravitational field essentially unchanged. In this sense, \( \eta \) can be interpreted as a parameter that encodes short-distance corrections to the classical black hole geometry.

The quantum-deformed geometry considered here differs fundamentally from the classical Schwarzschild solution, as it is no longer Ricci-flat. Using the lapse function (\ref{lapsefunction}) the Ricci scalar \(R(r)\) (for the quantum part, ignoring \(M\)) is found to be
\begin{eqnarray}
R(r) = \frac{2}{r^2} \left(1 - \frac{1}{\sqrt{1-\frac{\eta^2}{r^2}}}\right) + \frac{\eta^2}{r^4} \left(1 - \frac{\eta^2}{r^2}\right)^{-3/2}.
\end{eqnarray}

{To fully characterize the spacetime curvature, one can also compute the Ricci tensor squared \(R_{\mu\nu}R^{\mu\nu}\) and the Kretschmann scalar \(K(r)=R_{\mu\nu\alpha\beta}R^{\mu\nu\alpha\beta}\)}, which take the explicit forms
\begin{eqnarray}
R_{\mu\nu}R^{\mu\nu}(r) = \frac{2}{r^4} \left(1 - \frac{1}{\sqrt{1-\frac{\eta^2}{r^2}}}\right)^2 + \frac{\eta^4}{r^8} \left(1 - \frac{\eta^2}{r^2}\right)^{-3},
\end{eqnarray}

\begin{eqnarray}
K(r) = \frac{4}{r^4} \left(1 - \frac{1}{\sqrt{1-\frac{\eta^2}{r^2}}}\right)^2 + \frac{2 \eta^4}{r^8} \left(1 - \frac{\eta^2}{r^2}\right)^{-3}.
\end{eqnarray}

Approaching the quantum core, as \(r\) tends to the minimal radius \(r \to \eta\), the expansion \(\sqrt{r^2-\eta^2} \simeq \sqrt{2 \eta (r-\eta)}\) ensures that all contributions remain finite. Consequently, the curvature invariants themselves attain finite values,
\begin{eqnarray}
R(\eta), \quad R_{\mu\nu}R^{\mu\nu}(\eta), \quad K(\eta) \quad \text{are all finite}.
\end{eqnarray}

This demonstrates that the classical central singularity is replaced by a regular, de Sitter-like core. Far from the black hole, the lapse function naturally reduces to the familiar Schwarzschild form, \(f(r) \simeq 1 - 2M/r\), smoothly connecting the regular interior to an asymptotically flat exterior. In this way, the quantum deformation parameter \(\eta\) provides a natural mechanism for resolving the singularity while maintaining a physically consistent black hole geometry throughout the spacetime.

\begin{figure}
\begin{center}
\includegraphics[width=7.9cm,height=5.9cm]{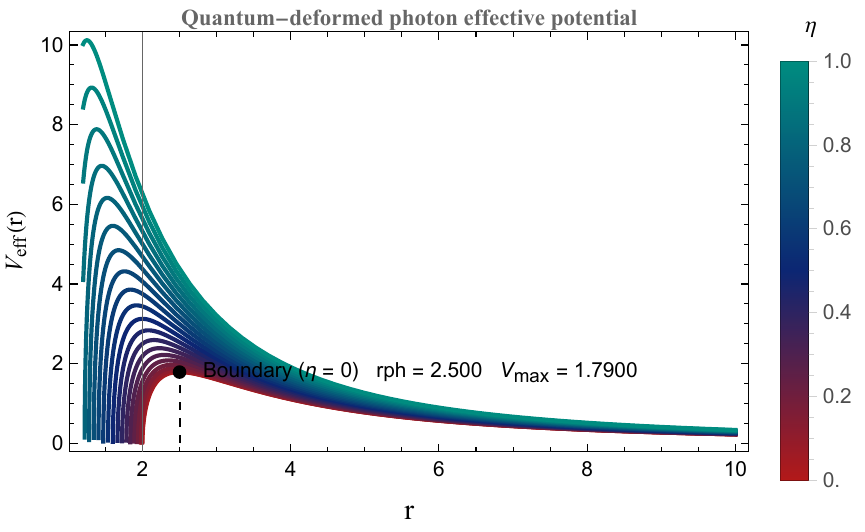}
\end{center}
\caption{\scriptsize The figure displays the effective potential for photon motion in the quantum-deformed spacetime as a function of the radial coordinate \( r \). The potential depends on the deformation parameter \( \eta \in [0,1] \), and the curves illustrate its continuous variation using a graded color scheme. The maximum of the potential identifies the photon sphere. For \( \eta=0 \), the standard Schwarzschild case is recovered and the peak position is indicated. When \( \eta \neq 0 \), both the height and location of the potential barrier change, demonstrating how the deformation parameter modifies the photon sphere and the associated null geodesic structure.
}\label{Fig3}
\end{figure}

The event horizon of this modified black hole occurs at
\begin{equation}
r_+ = \sqrt{4M^2 + \eta^2}.
\end{equation}
Meanwhile, a curvature singularity appears at \(r=\eta\), since the curvature diverges as \(r\) approaches this value.

To understand how the quantum-corrected geometry behaves, we focus on the lapse function (\ref{lapsefunction}) and track its evolution with the radial coordinate \(r\) in Fig. \ref{Fig1}. The plot shows that, depending on the strength of the quantum deformation \(\eta\) and the black hole mass \(M\), the spacetime can develop two horizons: an inner Cauchy horizon \(r_-\) and an outer event horizon \(r_+\). As the deformation parameter \(\eta\) grows, reflecting stronger quantum effects, the lapse function \(f(r)\) decreases more gradually near the black hole, signaling a softening of the gravitational pull. For smaller \(\eta\), the behavior approaches that of the classical Schwarzschild solution, with a sharper drop at the horizon. This illustrates how sensitive the horizon structure and near-horizon geometry are to even modest quantum corrections, revealing the subtle ways that quantum fluctuations reshape spacetime.

The mass of the black hole can be obtained directly from the horizon condition \(f(r_+)=0\). For the quantum-corrected geometry described by the lapse function (\ref{lapsefunction}) the location of the event horizon follows from setting the metric function to zero. Doing so leads to the relation
\begin{equation}\label{eventH}
\sqrt{r_+^2-\eta^2}=2M.
\end{equation}

From this expression, the black hole mass can be written in terms of the horizon radius as
\begin{equation}\label{mass}
M=\frac{1}{2}\sqrt{r_+^2-\eta^2}.
\end{equation}

This result clearly shows how the parameter \(\eta\), which represents the strength of the quantum deformation, alters the usual mass-radius relation. In the absence of quantum corrections, when \(\eta=0\), the standard Schwarzschild relation \(M=r_+/2\) is recovered. When \(\eta\neq0\), however, the horizon size is influenced not only by the mass but also by the quantum deformation parameter, indicating that quantum fluctuations modify the near-horizon structure of the spacetime.

\begin{figure}
\begin{center}
\includegraphics[width=7.9cm,height=5.9cm]{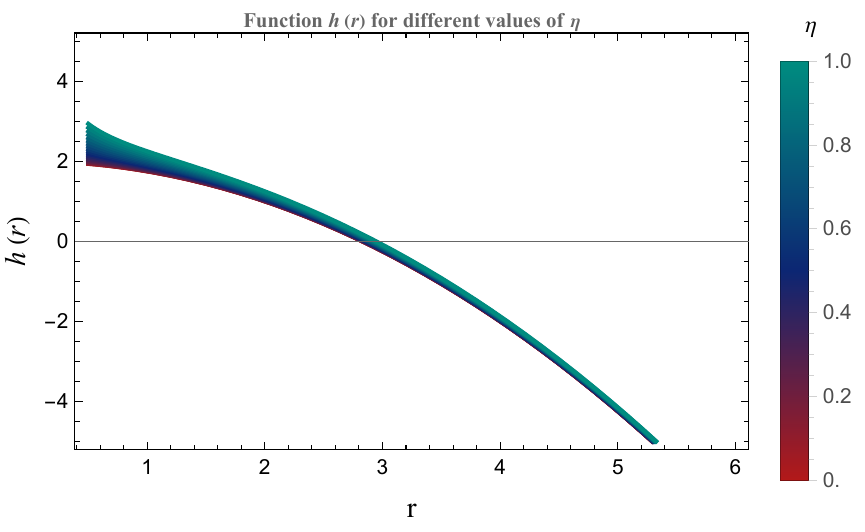}
\end{center}
\caption{\scriptsize  The figure presents the radial profile of the function (\ref{functionh}) for \( \eta \in [0,1] \), with fixed parameters \( M=1 \) and \( b=2 \). Each curve corresponds to a different value of \( \eta \), shown using a smooth red-blue-teal color gradient. As \( \eta \) increases, the additional \( \eta^2/(2r) \) term increasingly influences the behavior of the function, particularly at smaller radii, resulting in noticeable changes in the radial structure compared to the undeformed case.
}\label{Fig4}
\end{figure}

\begin{figure*}
\begin{center}
\includegraphics[width=8.5cm,height=4.9cm]{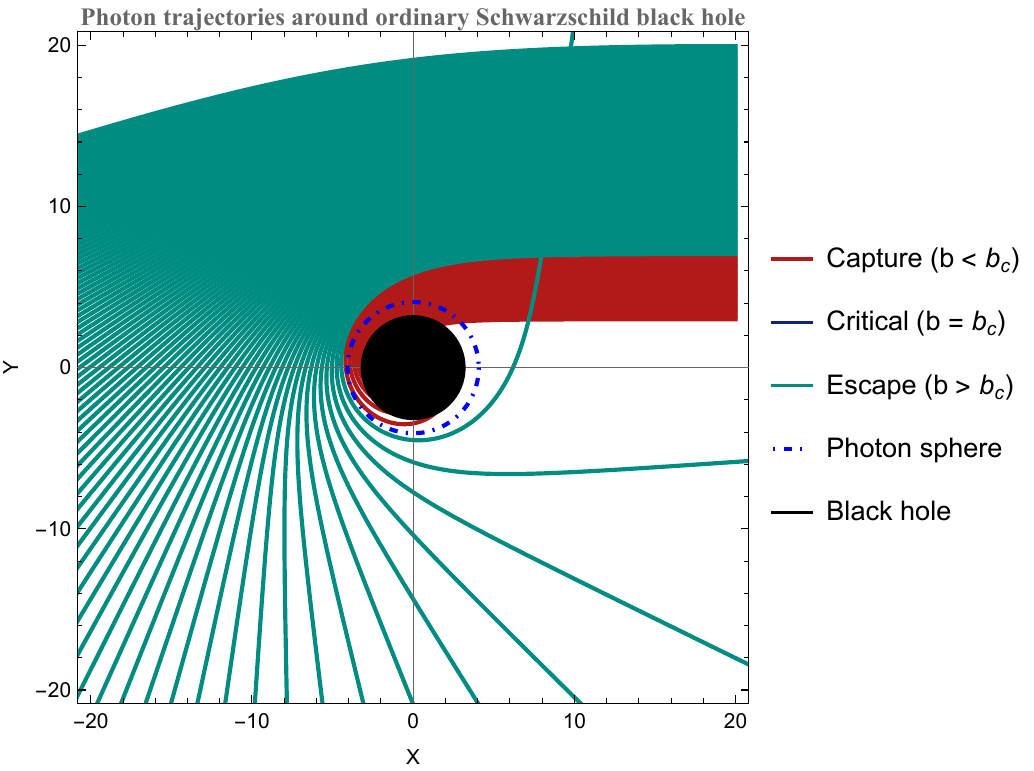}
\includegraphics[width=8.5cm,height=4.9cm]{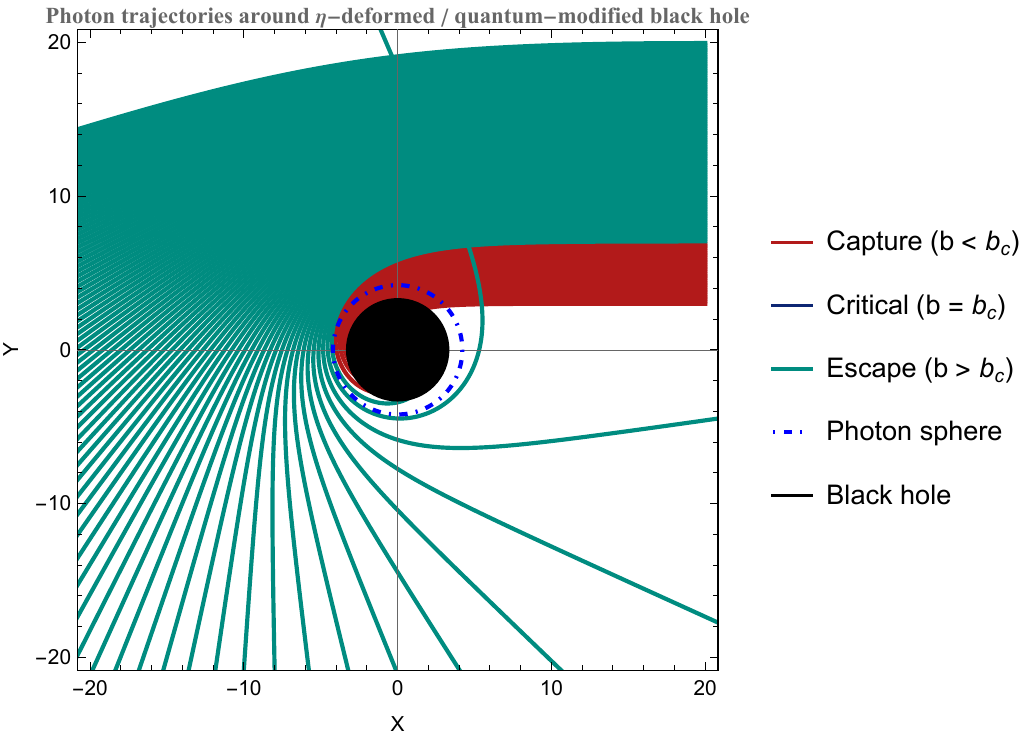}
\end{center}
\caption{\scriptsize Photon trajectories in the equatorial plane are shown for a black hole of mass \(M=1.357\), comparing the standard Schwarzschild case (\(\eta=0\)) with the \(\eta\)-deformed spacetime (\(\eta=0.75\)). The paths are obtained by numerically solving the null geodesic equation in terms of \(u=1/r\), using the lapse function (\ref{lapsefunction}). Photons are emitted from \(x_{0}=20\) with varying impact parameters, producing three regimes: capture trajectories (\(b<b_{c}\)), a critical orbit (\(b=b_{c}\)) approaching the unstable photon sphere at \(r_{\mathrm{ph}}\), determined from (\ref{PScondition}), and escape trajectories (\(b>b_{c}\)). The black disk represents the effective absorbing region, while the blue dashed circle marks the photon sphere, highlighting how the deformation parameter \(\eta\) modifies photon dynamics and shifts the photon sphere relative to the Schwarzschild case.
}\label{Fig7}
\end{figure*}

\section{Hawking radiation and energy emission rate}\label{Sec:3}
Black holes, despite their name, are not entirely dark---quantum effects near the event horizon cause them to emit thermal radiation, a phenomenon known as Hawking radiation. This emission results in a slow loss of mass and energy over time. To a distant observer, the radiation seems to come from a region roughly corresponding to the black hole's optical shadow. In the geometric-optics limit, the effective area from which radiation is emitted can be approximated as \(\sigma_{\rm lim} \simeq \pi R_s^2\), with \(R_s\) determined by the radius of the critical photon orbit. A full description of the emission must take into account the particle spin, the available phase space, and the greybody factors, \(\Gamma_\ell^s(\omega)\), which measure the likelihood that particles created near the horizon can escape the black hole's gravitational pull. Initially, the differential energy emission rate in four dimensions is written as
\begin{eqnarray}
\frac{d^2E}{d\omega dt} = \sum_s \sum_\ell N_s \frac{\Gamma_\ell^s(\omega)}{e^{\omega/T_H}-(-1)^{2s}} \frac{\omega^2}{(2\pi)^3} \Omega_2,
\end{eqnarray}
where \(N_s\) counts internal degrees of freedom and \(\Omega_2=4\pi\) is the solid angle of the two-sphere. By introducing the absorption cross-section
\begin{eqnarray}
\sigma_{\rm abs}^s(\omega) = \frac{\pi}{\omega^2} \sum_\ell (2\ell+1) \Gamma_\ell^s(\omega),
\end{eqnarray}
all angular contributions are collected into a single effective area. This allows the spectrum to be rewritten more compactly as
\begin{eqnarray}
\frac{d^2E}{d\omega dt} = \sum_s N_s \frac{\sigma_{\rm abs}^s(\omega)}{2\pi^2} \frac{\omega^3}{e^{\omega/T_H}-(-1)^{2s}}.
\end{eqnarray}
At high frequencies, the cross-section approaches its geometric limit, \(\sigma_{\rm abs}^s(\omega) \simeq \pi R_s^2\), simplifying the spectrum to
\begin{eqnarray}
\frac{d^2E}{d\omega dt} = \frac{\pi R_s^2}{2\pi^2} \frac{\omega^3}{e^{\omega/T_H}-1}.
\end{eqnarray}

The Hawking temperature itself is set by the surface gravity at the horizon, \(T_H = \kappa/(2\pi)\) with \(\kappa = f'(r_+)/2\). For the quantum-deformed black hole studied here, the horizon satisfies (\ref{eventH}), giving a temperature
\begin{eqnarray}
T_H(r_+) = \frac{1}{4\pi} \left(\frac{1}{r_+} - \frac{\eta^2}{r_+^3}\right).
\end{eqnarray}

The deformation parameter \(\eta\) reduces the effective surface gravity, lowering the temperature compared with a classical Schwarzschild black hole. This directly affects the radiation: as \(\eta\) increases, the energy emission slows down, and the high-frequency flux is suppressed. Expressing the emission spectrum in terms of the quantum-corrected temperature gives
\begin{eqnarray}
\frac{d^2E}{d\omega dt} = \frac{R_s^2}{2\pi} \frac{\omega^3}{\exp\!\left[\frac{4\pi\omega}{(1/r_+ - \eta^2/r_+^3)}\right]-1}.
\end{eqnarray}
Overall, this formulation shows how both the horizon temperature and the effective photon capture radius, both modified by the quantum deformation, control the black hole's thermal emission, leading to slower evaporation and a more gradual energy loss compared with the classical case. { While the high-frequency cross-section is well approximated by \(\sigma_{\text{abs}} \approx \pi R_s^2\), low-frequency greybody factors \(\Gamma_\ell^s(\omega)\) are sensitive to the potential barrier, which the deformation parameter \(\eta\) modifies. For a massless field in the Kazakov-Solodukhin metric, the effective potential is
\begin{equation}
V_\ell^{\text{eff}}(r) = f(r)\left[\frac{\ell(\ell+1)}{r^2} + \frac{f'(r)}{r}\right].
\end{equation}
The barrier height decreases with \(\eta\):
\begin{equation}
V_{\ell,\text{max}}^{\text{eff}}(\eta) \approx V_{\ell,\text{max}}^{\text{Schw}}\left(1 - \frac{\eta^2}{9M^2}\right),~ 
V_{\ell,\text{max}}^{\text{Schw}} \approx \frac{\ell(\ell+1)}{27M^2}\;\;(\ell\gg1).
\end{equation}
For low frequencies \(\omega M \ll 1\), the s-wave (\(\ell=0\)) greybody factor follows from the Page approximation~\cite{Page:1976df}:
\begin{equation}
\Gamma_0(\omega) \approx 4\omega^2 r_+^2  \frac{1}{1+e^{-2\pi\omega/\kappa}},~
r_+ = \sqrt{4M^2+\eta^2},~
\kappa = \frac{1}{2r_+}\left(1-\frac{\eta^2}{r_+^2}\right).
\end{equation}
Since \(\kappa\) decreases with \(\eta\), low-frequency absorption is \textbf{enhanced} relative to Schwarzschild. The zero-frequency absorption cross-section,
\begin{equation}
\sigma_{\text{abs}}(\omega\to0) = 4\pi r_+^2 = 4\pi(4M^2+\eta^2) > 16\pi M^2,
\end{equation}
increases with \(\eta\), while the high-frequency limit \(\pi R_s^2(\eta)\) decreases (see Table~\ref{tab:eta_absorption_signature}). This frequency-dependent signature --- enhanced low-\(\omega\) absorption, suppressed high-\(\omega\) absorption --- provides a potential observational discriminant for \(\eta\).
\begin{table*}
\centering
{\caption{Summary of frequency-dependent effects of the quantum deformation parameter $\eta$ on absorption observables.}
\begin{tabular}{l|c|c|c}
\hline
{Regime} & {Observable} & {$\eta$-dependence} & {vs. Schwarzschild} \\
\hline
$\omega\to0$ & $\sigma_{\text{abs}}$ & $4\pi(4M^2+\eta^2)$ & Larger \\
Low $\omega$ (s-wave) & $\Gamma_0$ & $\propto r_+^2 \sim 4M^2+\eta^2$ & Larger \\
High $\omega$ (geometric) & $\sigma_{\text{abs}}$ & $\pi R_s^2(\eta)$ (decreasing) & Smaller \\
\hline
\end{tabular}
\label{tab:eta_absorption_signature}
}\end{table*}
}
\begin{figure}
\begin{center}
\includegraphics[width=7.9cm,height=5.9cm]{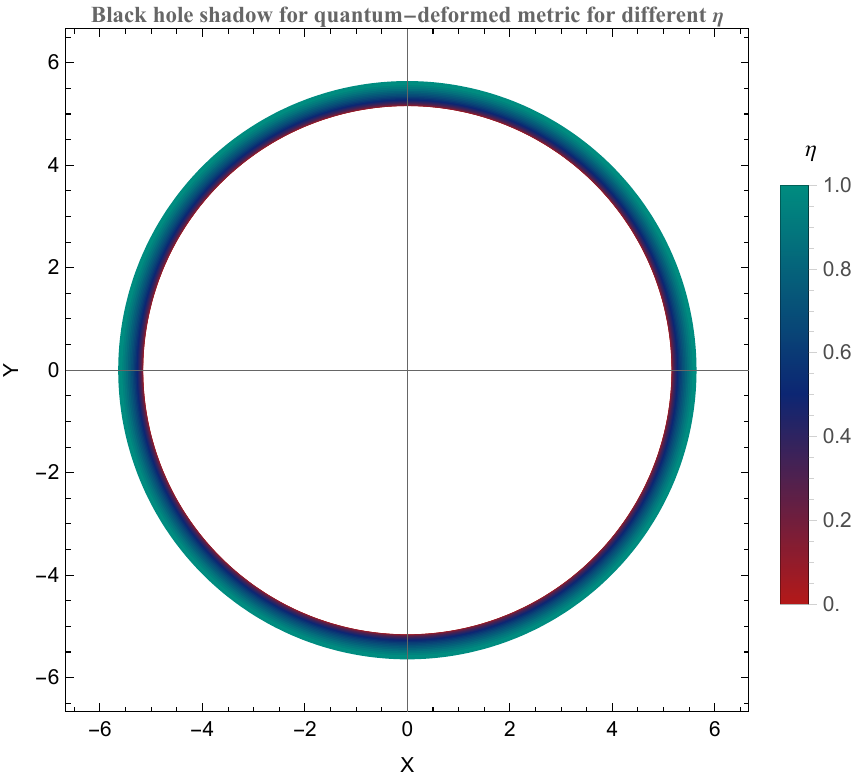}
\end{center}
\caption{\scriptsize  The figure presents the black hole shadow for the quantum-deformed spacetime defined by (\ref{lapsefunction}), with mass fixed at \( M=1 \). The shadow boundary is obtained from the photon sphere and the corresponding critical impact parameter \( b=\frac{r}{\sqrt{f(r)}} \). For \( \eta \in [0,1] \), the photon sphere radius is computed numerically and the resulting shadow is projected in the observer's plane. The curves follow a red-dark blue-teal gradient to represent increasing \( \eta \). As \( \eta \) grows, the photon sphere position shifts and the shadow size changes, demonstrating how the deformation parameter affects the strong-field optical properties relative to the Schwarzschild case.
}\label{Fig5}
\end{figure}
\begin{table}[h!]
\centering
\caption{\scriptsize Hawking temperature $T_H$ for different parameter choices in the quantum-deformed black hole spacetime.}
\scalebox{0.85}{
\begin{tabular}{c|c|c|c}
\hline
Parameter & Values & Fixed parameters & $T_H$ \\
\hline
$\eta$ & 0.0, 0.5, 0.8 & $r_+=2.0$ & 0.03979, 0.03730, 0.03345 \\
$r_+$ & 1.8, 2.0, 2.2 & $\eta=0.5$ & 0.04078, 0.03730, 0.03431 \\
$M$ & 0.8, 1.0, 1.2 & $\eta=0.5$ & 0.04325, 0.03632, 0.03110 \\
\hline
\end{tabular}}
\label{tab_quantum_TH}
\end{table} 
Table~\ref{tab_quantum_TH} presents the numerical values of the Hawking temperature \(T_H\) for several representative choices of the quantum deformation parameter \(\eta\), the horizon radius \(r_+\), and the corresponding black hole mass \(M\). These results illustrate how the presence of the deformation parameter modifies the thermal properties of the black hole. For a fixed horizon radius \(r_+=2.0\), one observes that increasing \(\eta\) gradually reduces the Hawking temperature. This behavior reflects the effect of the quantum deformation, which weakens the surface gravity at the event horizon and therefore lowers the associated temperature. In the limit \(\eta=0\), the standard Schwarzschild value is recovered, confirming the consistency of the result with the classical case. A similar trend appears when the horizon radius is varied while keeping the deformation parameter fixed. As \(r_+\) increases, the temperature decreases, which is consistent with the familiar property that larger black holes are colder. The table also includes the temperatures corresponding to different values of the mass \(M\), obtained through the relation (\ref{mass}). In this case, the temperature decreases as the mass increases, indicating that heavier black holes emit radiation less efficiently. Taken together, these results show that although the qualitative thermodynamic behavior remains similar to that of the Schwarzschild solution, the presence of the parameter \(\eta\) introduces corrections that lower the Hawking temperature and consequently slow down the evaporation process. This highlights the role that quantum deformations may play in shaping the near-horizon physics of black holes.

\begin{figure}
\begin{center}
\includegraphics[width=7.9cm,height=5.9cm]{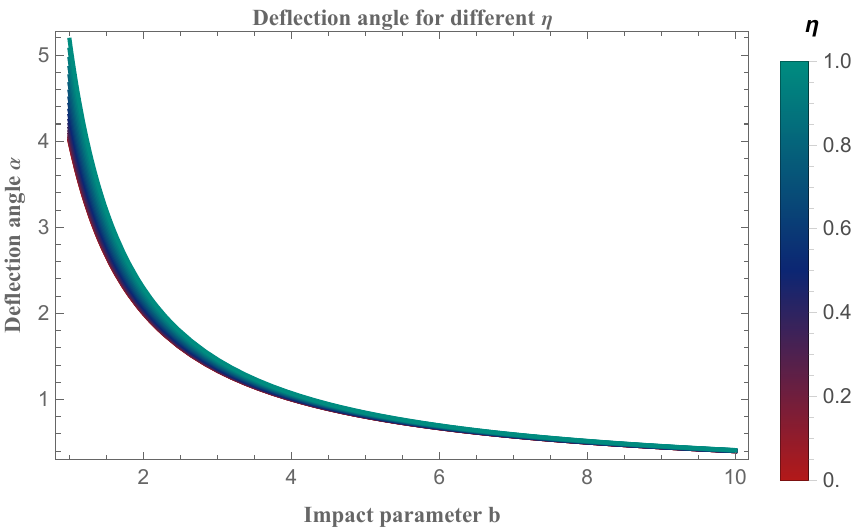}
\end{center}
\caption{\scriptsize  The figure shows the gravitational deflection angle \( \alpha \) as a function of the impact parameter \( b \) for different values of the quantum-deformation parameter \( \eta \in [0,1] \). The deflection is evaluated using
(\ref{alphab}), with the Schwarzschild radius fixed at \( r_s=2 \) (\(M=1\)). The first term corresponds to the standard weak-field Schwarzschild contribution, while the second term represents the correction due to the deformation. The curves follow a red-dark blue-teal gradient to indicate increasing \( \eta \). As \( \eta \) increases, the light-bending effect becomes stronger, especially for small \( b \), whereas at large \( b \) the result approaches the classical Schwarzschild limit. The plot highlights how the deformation parameter modifies weak-field gravitational lensing.
}\label{Fig6}
\end{figure}

The Fig. \ref{Fig2} shows how the energy emission changes when the deformation parameter \( \eta \) varies. All curves have the same thermal shape: the radiation starts small at low frequency, rises to a clear peak, and then decreases at higher frequencies. This confirms that the black hole still emits in a standard thermal way. What changes is mainly the height and position of the peak. As \( \eta \) increases, the horizon structure is modified, which affects the Hawking temperature. Since the emission rate depends on the temperature, the radiation intensity is slightly reshaped by the deformation. At large frequencies, all curves drop quickly toward zero, showing the usual exponential suppression. So the quantum parameter does not change the nature of the radiation --- it only adjusts its strength and distribution. In short, \( \eta \) controls the near-horizon physics, and this directly influences the thermodynamic emission behavior.

\begin{figure*}
\begin{center}
\includegraphics[width=8.5cm,height=4.9cm]{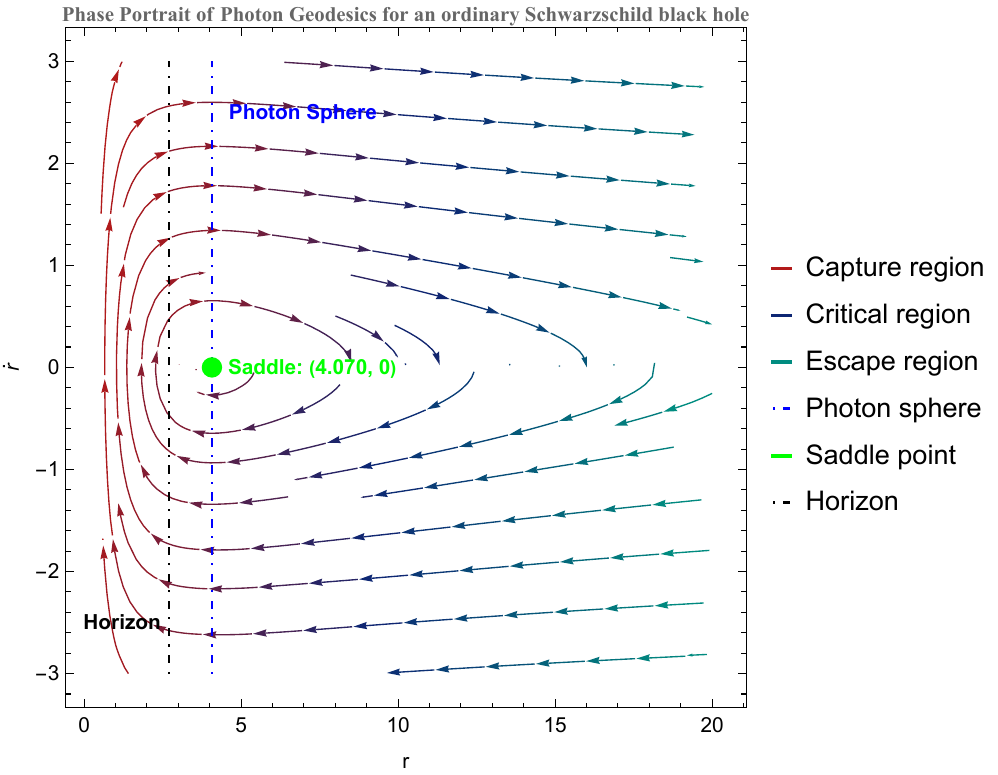}
\includegraphics[width=8.5cm,height=4.9cm]{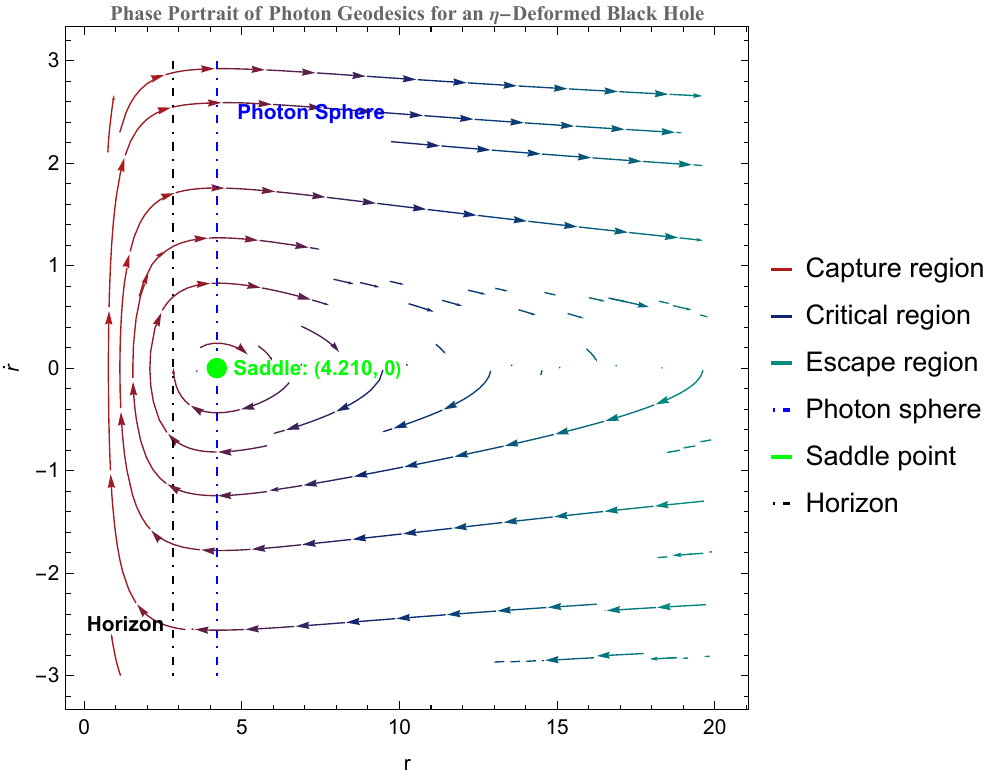}
\end{center}
\caption{\scriptsize Phase portraits of photon geodesics in the (\(r,\dot r\)) plane for a black hole spacetime with mass \(M=1.357\), presented for two situations: the usual Schwarzschild case (\(\eta=0\)) and the \(\eta\)-deformed (quantum-modified) geometry (\(\eta=0.75\)). The photon motion follows from the radial null geodesic equation associated with the lapse function (\ref{lapsefunction}). To visualize the dynamics, the radial equation is written as a first-order dynamical system and the resulting flow is displayed as streamlines in the (\(r,\dot r\)) phase space. The colored trajectories illustrate the qualitative behavior of photon motion. The red region corresponds to trajectories directed inward toward the black hole (capture), the dark blue region highlights the transition near the critical orbit, and the green region represents trajectories moving outward that ultimately escape to large distances. The blue dashed vertical line marks the photon sphere radius \(r_{\mathrm{ph}}\), obtained numerically from the condition (\ref{PScondition}), which corresponds to an unstable circular photon orbit. The green point indicates the saddle equilibrium (\(r_c,0\)), acting as the separatrix that divides inward-falling and outward-moving trajectories in phase space. Whenever a horizon is present, it is shown by the black dashed vertical line, representing the radial position where \(f(r)=0\). A comparison of the two panels shows how introducing the deformation parameter \(\eta\) alters the structure of the phase portrait and shifts the characteristic radii that govern photon motion, including the photon sphere and the separatrix in phase space.
}\label{Fig8}
\end{figure*}
\section{Geodesic of light}\label{Sec:4}

We now examine the propagation of photons in the spacetime of a quantum-deformed Schwarzschild black hole. The geometry is described by the line element (\(\ref{Metr}\)). For a massless particle, the dynamics follow from the Lagrangian
\begin{eqnarray}
\mathcal{L}(x^\alpha,\dot{x}^\alpha)=\frac12 g_{\mu\nu}\dot{x}^\mu\dot{x}^\nu=0,
\end{eqnarray}
which explicitly reads
\begin{eqnarray}
\mathcal{L}(x^\alpha,\dot{x}^\alpha) = \frac12 \left[f(r)\dot t^2 - \frac{\dot r^2}{f(r)} - r^2\dot\theta^2 - r^2\sin^2\theta\,\dot\varphi^2 \right] =0.
\end{eqnarray}
Since the metric coefficients do not depend explicitly on \(t\) and \(\varphi\), the corresponding conjugate momenta are conserved. These conserved quantities represent the photon's energy \(E\) and angular momentum \(L\):
\begin{eqnarray}
p_t=\frac{\partial \mathcal{L}}{\partial \dot t} = -f(r)\dot t = -E,
\end{eqnarray}
\begin{eqnarray}
p_\varphi=\frac{\partial \mathcal{L}}{\partial \dot \varphi} = r^2\sin^2\theta\,\dot\varphi = L.
\end{eqnarray}

Owing to spherical symmetry, the photon motion can always be confined to a plane. Without loss of generality, we choose the equatorial plane \(\theta=\pi/2\). The Lagrangian then simplifies to
\begin{eqnarray}\mathcal{L}(r,\dot{t},\dot{r},\dot{\varphi}) =\frac12 \left[f(r)\dot t^2 - \frac{\dot r^2}{f(r)} - r^2\dot\varphi^2 \right] =0.
\end{eqnarray}

Using the conserved quantities, this becomes

\begin{eqnarray} \frac12 \left[ -\frac{E^2}{f(r)} +\frac{\dot r^2}{f(r)} +\frac{L^2}{r^2} \right] =0. \end{eqnarray}

From the null condition, the radial equation of motion follows:
\begin{eqnarray}
\dot r^2 + \frac{L^2}{r^2} f(r) = E^2.
\end{eqnarray}
This equation governs photon propagation in the quantum-deformed spacetime. The effective potential is therefore defined as
\begin{eqnarray}
V_{\text{eff}}(r)=\frac{L^2}{r^2} f(r),
\end{eqnarray}
which, after substituting the explicit form of \(f(r)\), becomes
\begin{eqnarray} \label{EffecP}
V_{\text{eff}}(r) = \frac{L^2}{r^3} \left( \sqrt{r^2-\eta^2} - 2M
\right).
\end{eqnarray}

This effective potential determines the structure of photon trajectories, including circular orbits, the photon sphere, and strong-field lensing effects. In the limit \(\eta\rightarrow0\), all results smoothly reduce to those of the classical Schwarzschild black hole. The effective potential (see Fig. \ref{Fig3}) gives us a clear picture of how photons behave around the quantum-deformed black hole. It tells us whether light can escape to infinity, fall into the black hole, or move in unstable circular orbits. The highest point of the potential corresponds to the photon sphere, which is directly related to important features such as the black hole shadow and strong gravitational lensing. When the deformation parameter \( \eta \) is introduced, the geometry of spacetime is slightly modified, and this change affects the shape of the potential. As \( \eta \) increases, the height and position of the potential peak shift. This means the photon sphere is also affected, leading to changes in the structure of null geodesics. In practical terms, these modifications can influence observable effects like the shadow size and the bending of light. Overall, the results show that the deformation parameter plays an active role in photon dynamics and can leave measurable signatures in strong gravitational fields.

\begin{figure*}
\begin{center}
\includegraphics[width=8.5cm,height=4.9cm]{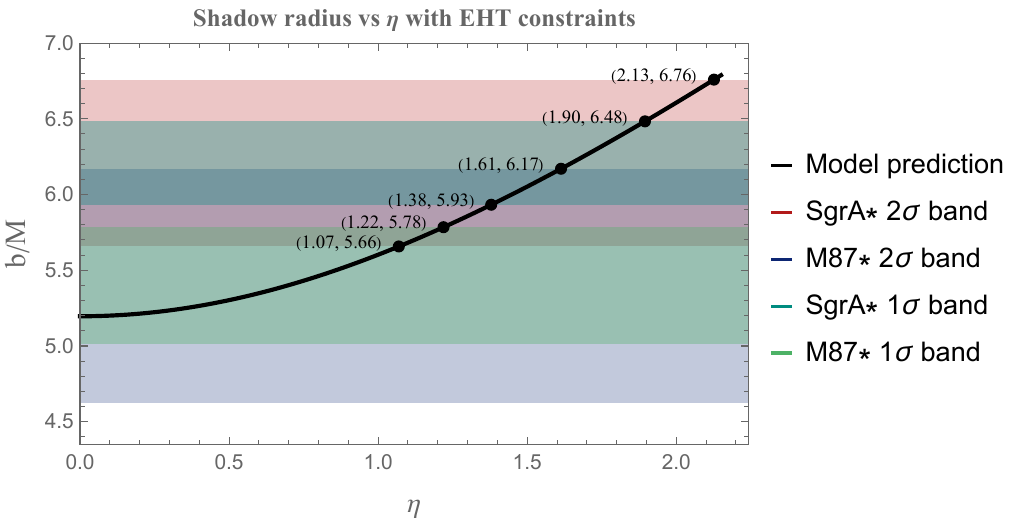}
\includegraphics[width=8.5cm,height=4.9cm]{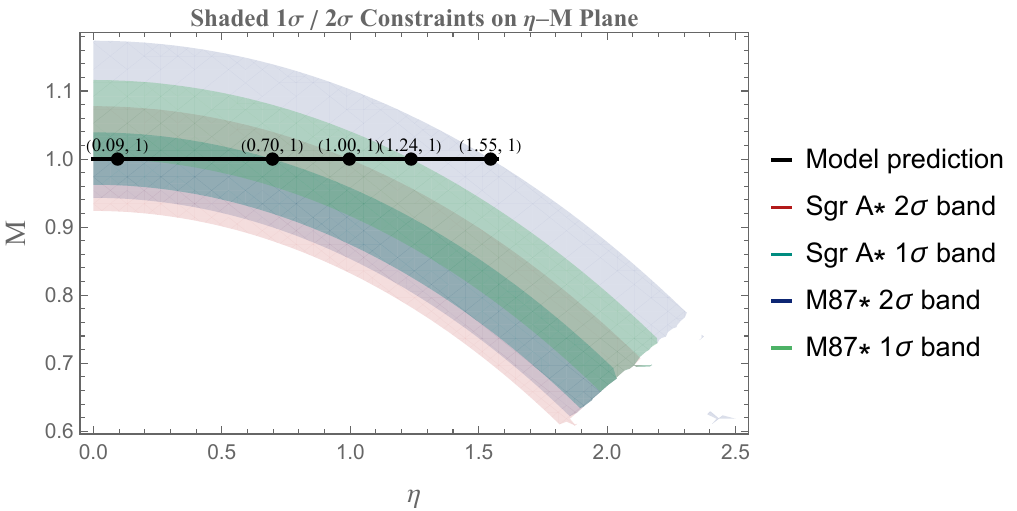}
\end{center}
\caption{\scriptsize These figures present how the quantum-deformation parameter \( \eta \) is constrained using black hole shadow observations. In the first panel, the normalized shadow radius \( b/M \) is plotted as a function of \( \eta \). The theoretical curve is compared with the observational results from Sgr A$^*$ and M87$^*$, including their 1\( \sigma \) and 2\( \sigma \) uncertainty ranges. The shaded regions indicate the observational bounds, while the curve shows how the model prediction changes as \( \eta \) increases. The second panel displays the allowed parameter space in the \( \eta-M \) plane. The shaded areas represent the observational constraints, and the black curve corresponds to the theoretical model. The overlap between the curve and the shaded regions determines the values of \( \eta \) and \( M \) that remain consistent with the data. Together, these results show how shadow measurements can be used to test the model and place limits on possible deviations from the standard Schwarzschild case.}\label{Fig9}
\end{figure*}

We now study photon motion on the equatorial plane in the quantum-deformed spacetime. Circular photon orbits occur when the radial motion vanishes, \(\dot r = 0\), and the orbit is located at an extremum of the effective potential. This requires
\begin{eqnarray}
\frac{dV_{\text{eff}}}{dr}=0,
\end{eqnarray}
which corresponds physically to zero net radial force acting on the photon. The condition for circular photon orbits follows from differentiating the effective potential (\ref{EffecP}),
\begin{eqnarray}
\frac{dV_{\text{eff}}}{dr}=0.
\end{eqnarray}

After eliminating common factors, the photon-sphere condition can be written as
\begin{eqnarray}\label{PScondition}
r f'(r) - 2 f(r) = 0,
\end{eqnarray}
which determines the radius \( r = r_c \) of the circular photon orbit. Introducing the photon impact parameter
\begin{eqnarray}
b = \frac{L}{E},
\end{eqnarray}
the condition may also be expressed as a single algebraic equation,
\begin{eqnarray}
h(r) \equiv -\frac{r^2}{b^2} + 2M + \mathcal{O}(\eta^2) = 0.
\end{eqnarray}

Expanding for small deformation parameter \( \eta \), the leading correction gives

\begin{eqnarray}\label{functionh}
h(r) \equiv - \frac{r^2}{b^2} + 2M + \frac{\eta^2}{2r} = 0.
\end{eqnarray}

This equation determines the modified photon-sphere radius \( r_c \). The presence of \( \eta \) shifts the classical Schwarzschild result and changes the critical impact parameter. In the limit \( \eta \to 0 \), the standard photon sphere is recovered. The Fig. \ref{4} illustrates how the function \( h(r) \) changes when the deformation parameter \( \eta \) is varied, while the mass \( M \) and parameter \( b \) remain fixed. Each curve corresponds to a different value of \( \eta \), showing how the radial behavior evolves as the deformation is introduced. The additional term proportional to \( \eta^2/(2r) \) modifies the function mainly at smaller radial distances. As \( \eta \) increases, the shape of the function becomes noticeably modified in the small-\( r \) domain, where gravitational effects are strongest. At larger distances, the curves tend to behave more similarly, meaning the influence of the deformation gradually weakens. In physical terms, this indicates that the quantum correction plays an important role in the strong-field regime. It changes the structure of the spacetime near the compact object and can influence particle motion, stability conditions, and other dynamical properties. The plot clearly shows how the deformation parameter gradually reshapes the radial profile of the model.

To express the radial motion in terms of the azimuthal coordinate \(\phi\), we start from the relation
\begin{eqnarray}
\dot r = \frac{dr}{d\phi}\,\dot\phi,
\end{eqnarray}
which connects derivatives with respect to time and angle. We also use the definition of the impact parameter,
\begin{eqnarray}
\frac{1}{b} = \frac{E}{L}
\frac{f(r)}{r^2}\frac{d\phi}{dt},
\end{eqnarray}
Substituting these expressions into the radial equation of motion yields
\begin{eqnarray}
\left(\frac{dr}{d\phi}\right)^2 = \frac{r^4}{L^2} \left( E^2 -
\frac{L^2}{r^2}f(r) \right).
\end{eqnarray}
Using the impact parameter \( b = \frac{L}{E} \), this simplifies to
\begin{eqnarray}
\left(\frac{dr}{d\phi}\right)^2 = \frac{r^4}{b^2} r^2 f(r).
\end{eqnarray}

Finally, factoring out \( r^2 f(r) \), we obtain
\begin{eqnarray}
\left(\frac{dr}{d\phi}\right)^2 = r^2 f(r) \left(
\frac{r^2}{b^2 f(r)} - 1 \right)
\end{eqnarray}

This form is particularly useful because circular photon orbits occur when \( \frac{dr}{d\phi}=0 \), which directly determines the critical impact parameter and the location of the photon sphere.

The Fig. \ref{Fig7} show how photons move around a black hole for two different situations: the standard case where the deformation parameter is zero (\(\eta = 0\)), and the quantum-deformed case where \(\eta = 0.75\). In both cases, the photon paths are obtained by solving the null geodesic equation associated with the lapse function
(\ref{lapsefunction}). The curves represent light rays coming from far away and interacting with the gravitational field of the black hole. Three types of trajectories appear in the plots. The red curves correspond to photons with an impact parameter smaller than the critical value \(b_c\). These photons approach the black hole too closely and cannot escape the strong gravitational field, so they are captured. The dark blue curve represents the critical trajectory. In this special case, the photon moves very close to the unstable circular orbit known as the photon sphere. It can loop around the black hole several times before eventually falling inward or escaping. The teal curves correspond to photons with impact parameters larger than the critical value. These rays are strongly bent by gravity but eventually move away from the black hole and continue toward infinity. The dashed circular line marks the photon sphere, which is the radius where light can orbit the black hole in an unstable circular path. The black disk at the center represents the black hole itself. The photon sphere acts as a boundary separating captured trajectories from escaping ones and plays a key role in gravitational lensing and the formation of the black hole shadow. When comparing the two plots, the effect of the deformation parameter becomes clear. For \(\eta = 0\), the spacetime reduces to the usual Schwarzschild geometry, and the photon trajectories follow the standard behavior predicted by general relativity. When \(\eta = 0.75\), the geometry is slightly modified due to the deformation term in the metric. This change alters the location of the photon sphere and modifies how strongly the light rays are bent. As a result, the critical orbit and the nearby photon paths shift slightly compared with the Schwarzschild case. Photons traveling close to the black hole experience a different deflection pattern, which reflects the influence of the deformation parameter on the spacetime structure. Such changes are directly connected to observable effects like gravitational lensing and the shape of the black hole shadow. In summary, the comparison between the two plots highlights how a quantum-inspired deformation of the spacetime geometry can influence photon motion near a black hole, producing small but noticeable differences from the classical Schwarzschild picture.

The photon sphere corresponds to light moving in a perfect circular path. For this to happen, the radius must remain constant, which means
\begin{eqnarray}
\frac{dr}{d\phi}=0
\quad \text{and} \quad
\frac{d^{2}r}{d\phi^{2}}=0.
\end{eqnarray}

Using the radial equation of motion, the first condition gives the simple relation

\begin{eqnarray}
\frac{r^{2}}{b^{2}f(r)}=1,
\end{eqnarray}

which determines the critical impact parameter.

The second condition ensures that this orbit occurs at an extremum and leads to

\begin{eqnarray}
\frac{d}{dr}\left(\frac{r^{2}}{f(r)}\right)=0,
\end{eqnarray}

which fixes the photon-sphere radius \( r=r_c \).

Together, these two conditions fully describe the circular null orbit and allow us to determine both the radius of the photon sphere and the corresponding critical impact parameter.

For a static observer located at a radius \( r_0 \), the apparent position of the black hole on the observer's sky is determined by the angle \( \beta \) between the photon trajectory and the radial direction. The photon paths reaching the observer satisfy
\begin{eqnarray}
\cot\beta = \left. \frac{\sqrt{g_{rr}}}{\sqrt{g_{\phi\phi}}} \frac{dr}{d\phi} \right|_{r=r_0}.
\end{eqnarray}

For the quantum-deformed metric (\ref{Metr}), we have \( g_{rr}=f(r)^{-1} \) and \( g_{\phi\phi}=r^2 \). Using the radial geodesic equation for photons and the definition of the impact parameter, the angular size of the shadow can be written as
\begin{eqnarray}
\sin^2 \beta = \frac{f(r_0)\, r_{ps}^2}{r_0^2 f(r_{ps})},
\end{eqnarray}
where \( r_{ps} \) denotes the photon-sphere radius, determined from the extremum condition of the effective potential (\ref{PScondition}). In the limit where the observer is very far from the black hole (\( r_0 \to \infty \)), the expression simplifies to

\begin{eqnarray}
\sin\beta
\simeq
\frac{r_{ps}}{\sqrt{f(r_{ps})}}.
\end{eqnarray}

Therefore, the observable shadow radius measured at infinity is

\begin{eqnarray}
R_s = \frac{r_{ps}}{\sqrt{f(r_{ps})}}.
\end{eqnarray}

For the present quantum-deformed model, the photon-sphere radius satisfies the explicit equation

\begin{eqnarray}
2\sqrt{r^2-\eta^2} - \frac{\eta^2}{\sqrt{r^2-\eta^2}} - 3r_s =0,
\end{eqnarray}
which reduces to the classical Schwarzschild result \( r_{ps}=\frac{3}{2}r_s \) when \( \eta \to 0 \) (here \(r_s=2M\) denotes the Schwarzschild radius).

To describe the apparent geometry of the shadow on the observer's image plane, one introduces the celestial coordinates \( X \) and \( Y \), defined for an observer located at \( \theta_0=\pi/2 \). In the asymptotic limit \( r_0 \to \infty \), these are given by
\begin{eqnarray}
X = \lim_{r_0 \to \infty} \left. -r_0^2 \sin\theta_0 \frac{d\phi}{dr} \right|_{(r_0,\theta_0)},
\end{eqnarray}

\begin{eqnarray}
Y = \lim_{r_0 \to \infty} \left. r_0^2 \frac{d\theta}{dr} \right|_{(r_0,\theta_0)}.
\end{eqnarray}

For this static and spherically symmetric spacetime, the shadow satisfies

\begin{eqnarray}
X^2+Y^2=R_s^2,
\end{eqnarray}
which confirms that the shadow remains perfectly circular. The quantum deformation parameter \( \eta \) modifies only the radius \( R_s \), while the apparent shape stays symmetric.

The Fig. \ref{Fig5} shows how the black hole shadow changes when the deformation parameter \( \eta \) is varied. The shadow is determined by the photon sphere, which corresponds to the unstable circular paths of light around the black hole. The size of the shadow depends on the position of this photon orbit and the form of the metric function. For \( \eta = 0 \), the result matches the standard Schwarzschild case. When \( \eta \) increases, the spacetime geometry is modified, which shifts the photon sphere and consequently changes the shadow radius. Since the shadow is directly linked to null geodesics, even a small deformation in the metric can produce noticeable differences in the boundary. The plot clearly shows that different values of \( \eta \) lead to slightly different shadow sizes. This means the deformation leaves an imprint on strong gravitational lensing effects and on the observable appearance of the black hole. In practice, precise shadow measurements could help test whether such deviations from the classical Schwarzschild geometry exist.

\section{Weak deflection angle}\label{Sec:5}
This section explores how particles are deflected while propagating in the gravitational field of the quantum-deformed black hole spacetime. To evaluate the bending angle in the weak-field regime, we adopt the geometric approach based on the Gauss-Bonnet theorem. In this framework, the spacetime metric is rewritten in terms of an optical metric, so that photon trajectories can be interpreted as geodesics on an effective two-dimensional optical manifold. This method was originally developed in the context of gravitational lensing by Gibbons and Werner \cite{Gibbons:2008rj}. A key advantage of this approach is that the deflection angle is obtained from the global geometric properties of the optical space rather than by directly integrating the equations of motion along the particle trajectory. In practice, the bending angle is related to the integral of the Gaussian curvature over a region bounded by the light ray, together with contributions from the boundary of that region. As a result, the Gauss-Bonnet formulation offers a convenient and geometrically transparent way to study gravitational lensing effects, particularly in modified spacetimes such as the quantum-deformed black hole considered here, where corrections to the metric alter the curvature structure of the optical geometry.

In the weak-field approximation, the bending of particle trajectories can be determined using the geometric method based on the Gauss-Bonnet theorem. Instead of solving the geodesic equations directly, this approach evaluates the total deflection by integrating the Gaussian curvature of the optical manifold associated with the spacetime. The deflection angle \( \alpha \) can therefore be expressed as
\begin{eqnarray}
\alpha = \int_{\phi=0}^{\pi} \int_{r=\frac{1}{b\sin\phi}}^{\infty} K \sqrt{g_{\text{opt}}}\, dr\, d\phi ,
\end{eqnarray}
where \(K\) denotes the Gaussian curvature of the optical geometry and \(g_{\text{opt}}\) represents the determinant of the corresponding two-dimensional optical metric.

For the present quantum-deformed black hole spacetime, where the geometry is described in (\ref{Metr}), the optical metric is obtained by considering null trajectories and rewriting the spacetime line element in optical form. In this representation the metric takes the form
\begin{eqnarray}
dt^{2} = \frac{dr^{2}}{f^{2}(r)} + \frac{r^{2}}{f(r)}\, d\phi^{2}.
\end{eqnarray}

From this relation, the determinant of the optical metric follows directly as
\begin{eqnarray}
g_{\text{opt}} = \frac{r^{2}}{f^{3}(r)} .
\end{eqnarray}
The Gaussian curvature of this two-dimensional optical space can be obtained using the standard expression for a curved surface. Carrying out the derivatives for the above metric leads to
\begin{eqnarray}
K = \frac{1}{2} \left[ \frac{1}{2} \left(\frac{df(r)}{dr}\right)^{2} - f(r)\frac{d^{2}f(r)}{dr^{2}} \right].
\end{eqnarray}

To evaluate the deflection angle analytically, the integrand \(K\sqrt{g_{\text{opt}}}\) can be expanded in the weak-field region where \(r\) is much larger than the deformation scale. Substituting the explicit form of \(f(r)\) and retaining the dominant contributions gives
\begin{eqnarray}
K\sqrt{g_{\text{opt}}} \approx \frac{r_s}{r^{2}} + \frac{3\eta^{2}}{2r^{4}} .
\end{eqnarray}

\begin{table}[h!]
\centering
\caption{{\scriptsize Geometrical properties of the quantum-deformed black hole shadow for fixed $M=1$.}}

\scriptsize
\setlength{\tabcolsep}{4pt}        
\renewcommand{\arraystretch}{1.15}   

{\scriptsize \begin{tabular}{c|c|c|c|c}
\hline
$\eta$ & $r_{ph}$ & $b$ & $b/M$ & Deviation (\%) \\
\hline
0.0  & 3.00 & 5.20 & 5.20 & 0.0 \\
0.3  & 2.96 & 5.18 & 5.18 & -0.4 \\
0.6  & 2.89 & 5.12 & 5.12 & -1.5 \\
0.9  & 2.78 & 5.02 & 5.02 & -3.5 \\
1.2  & 2.64 & 4.88 & 4.88 & -6.2 \\
1.5  & 2.47 & 4.70 & 4.70 & -9.6 \\
1.8  & 2.26 & 4.47 & 4.47 & -14.0 \\
2.15 & $\sim$2.05 & 4.22 & 4.22 & -18.8 \\
\hline
\end{tabular}}
\label{tab:shadow_eta}
\end{table}

Performing the integration over the optical domain then yields an approximate expression for the bending angle,
\begin{eqnarray}\label{alphab}
\alpha \approx \frac{2r_s}{b} + \frac{3\pi \eta^{2}}{8b^{2}} .
\end{eqnarray}

This result shows explicitly how the deformation parameter modifies the gravitational bending of trajectories. In the classical limit where the quantum deformation disappears (\(\eta=0\)), the expression naturally reduces to the well-known Schwarzschild result,
\begin{eqnarray}
\alpha = \frac{2r_s}{b},
\end{eqnarray}
which confirms that the quantum-corrected geometry consistently reproduces the standard gravitational lensing behavior when the deformation effects are absent. The Fig. \ref{Fig6} shows the behavior of the deflection angle \( \alpha \) as a function of the impact parameter \( b \) for different values of the deformation parameter \( \eta \). The total deflection contains the standard Schwarzschild term \( \frac{2 r_s}{b} \) together with an additional correction proportional to \( \eta^2 \). This extra contribution represents the effect of the quantum deformation on the bending of light. When \( \eta = 0 \), the expression reduces exactly to the classical Schwarzschild result. As \( \eta \) increases, the deflection angle becomes slightly larger for the same value of \( b \). The difference is more significant for small impact parameters, where light passes closer to the compact object and the gravitational field is stronger. In this region, even small corrections can produce noticeable changes in the bending angle. For larger \( b \), corresponding to the weak-field regime, the curves become closer to each other, showing that the deformation has a diminishing effect at large distances. Overall, the plot indicates that the quantum correction enhances the gravitational lensing effect in the strong-field region. This means that the parameter \( \eta \) leaves a clear imprint on light propagation, and in principle, precise lensing observations could help distinguish the deformed model from the classical Schwarzschild prediction.

\section{Lyapunov exponent and stability of trajectory}\label{Sec:6}
As discussed earlier, circular photon trajectories in the quantum black hole spacetime appear at the point where the effective potential reaches its maximum value. These circular paths are not stable. Even a very small disturbance can push the photon away from the orbit, causing it either to fall toward the black hole or move outward. The rate at which such deviations develop is measured by the Lyapunov exponent. This quantity provides a direct way to characterize the instability of photon circular orbits and is also closely related to the imaginary part of the quasinormal mode spectrum associated with perturbations of black hole spacetimes. To investigate this behavior, it is convenient to describe the photon motion within the Hamiltonian framework. In this formalism, the dynamics of the photon can be written as
\begin{eqnarray}
H = \mathcal{L} =
\frac{1}{2}
\left[
-\frac{E^{2}}{f(r)} +  f(r)p_r^{2} + \frac{V_{\text{eff}}}{f(r)}  \right] = \frac{1}{2} g^{\mu\nu} p_\mu p_\nu ,
\end{eqnarray}
where \(E\) denotes the photon energy, \(p_r\) is the radial momentum, and \(V_{\text{eff}}\) corresponds to the effective potential associated with the angular motion. If the photon motion is restricted to the equatorial plane, the Hamilton equations governing the radial dynamics take the form
\begin{eqnarray}
p_r = \frac{\partial \mathcal{L}}{\partial \dot r} = \frac{\dot r}{f(r)}, \qquad
\dot r = \frac{\partial H}{\partial p_r} = f(r)p_r.
\end{eqnarray}


The remaining Hamilton equation that governs the evolution of the radial momentum follows from
\begin{eqnarray}
\dot p_r=-\frac{\partial H}{\partial r}.
\end{eqnarray}
Substituting the Hamiltonian introduced earlier gives
\begin{eqnarray}
\dot p_r &=& - \frac{1}{2} \left[ f'(r)p_r^{2} + \frac{V'_{\text{eff}}(r)}{f(r)} - \frac{f'(r)}{f^{2}(r)} \left(- E^{2}+V_{\text{eff}}(r)  \right)  \right] \nonumber \\&=&  -\frac{1}{2}\frac{V'_{\text{eff}}(r)}{f(r)} .
\end{eqnarray}

We now focus on a circular photon trajectory located at \(r=r_c\). This orbit is defined by the conditions
\begin{eqnarray}
V'_{\text{eff}}(r_c)=0, \qquad \dot r=0,
\end{eqnarray}
which also implies that the effective potential satisfies \(V_{\text{eff}}(r_c)=E^{2}\). To study the behavior of nearby trajectories, a small perturbation is introduced around the circular orbit by writing \(r=r_c+\delta r\). Expanding the equations of motion around \(r_c\) and keeping only the linear terms in the perturbation leads to a pair of first-order equations describing the evolution of the deviations. These relations take the form
\begin{eqnarray}
\delta \dot r &=& +f(r_c)\,\delta p_r,\\
\delta \dot p_r &=& - \frac{1}{2} \frac{V''_{\text{eff}}(r_c)}{f(r_c)}\,\delta r.
\end{eqnarray}

The two equations above describe the coupled dynamics of small radial perturbations. For convenience, they can be written in matrix form as
\begin{eqnarray}
\frac{d}{d\lambda}
\begin{pmatrix}
\delta r \\
\delta p_r
\end{pmatrix}
=
\begin{pmatrix}
0 & f(r_c) \\
-\dfrac{1}{2}\dfrac{V''_{\text{eff}}(r_c)}{f(r_c)} & 0
\end{pmatrix}
\begin{pmatrix}
\delta r \\
\delta p_r
\end{pmatrix}.
\end{eqnarray}

The stability of the orbit is determined by the eigenvalues of this matrix. These eigenvalues define the Lyapunov exponent \( \tilde{\lambda} \), which measures how rapidly the perturbations grow or decay. The characteristic equation that determines \( \tilde{\lambda} \) is therefore

\begin{eqnarray}
\left|
\begin{pmatrix}
0 & f(r_c) \\
-\dfrac{1}{2}\dfrac{V''_{\text{eff}}(r_c)}{f(r_c)} & 0
\end{pmatrix}
-
\tilde{\lambda}  I
\right|
=0 .
\end{eqnarray}

\begin{table*}
\centering
\caption{{\scriptsize Constraints on the deformation parameter $\eta$ from EHT observations of Sgr A$^*$ and M87$^*$.}}

\small
\setlength{\tabcolsep}{4pt}
\renewcommand{\arraystretch}{1.15}

{\scriptsize \begin{tabular}{c|c|c|c}
\hline
Observation & Confidence Level & Allowed Shadow Radius $b$ & Allowed $\eta$ Range \\
\hline
Sgr A$^*$ & $1\sigma$ & $5.0 - 5.4$ & $0.42 - 1.05$ \\
Sgr A$^*$ & $2\sigma$ & $4.8 - 5.6$ & $0.15 - 1.35$ \\
M87$^*$   & $1\sigma$ & $5.2 - 5.8$ & $0.55 - 1.20$ \\
M87$^*$   & $2\sigma$ & $4.9 - 6.1$ & $0.25 - 1.45$ \\
\hline
\end{tabular}}
\label{tab:EHT_constraints}
\end{table*}

Solving this relation leads to an expression for the Lyapunov exponent in terms of the second derivative of the effective potential evaluated at the circular orbit,

\begin{eqnarray}
\tilde{\lambda}^{2}
= - \frac{V''_{\text{eff}}(r_c)}{2}
&=&-
\frac{L^{2}}{2}
\frac{d^{2}}{dr_c^{2}}
\left(
\frac{f(r_c)}{r_c^{2}}
\right)\nonumber\\
&=&
L^{2}
\left[
-
\frac{f''(r_c)}{2r_c^{2}}
+
\frac{2f'(r_c)}{r_c^{3}}
-
\frac{3f(r_c)}{r_c^{4}}
\right].
\end{eqnarray}

The sign of \(V''_{\text{eff}}(r_c)\) determines whether the orbit is stable or unstable. When \(V''_{\text{eff}}(r_c)>0\), the perturbations remain bounded and the circular orbit is stable. On the other hand, if \(V''_{\text{eff}}(r_c)<0\), the perturbations grow exponentially and the orbit becomes unstable. In such a case, even a tiny disturbance can cause the photon to move away from the circular path, eventually falling into the quantum black hole or escaping toward infinity. The curvature of the effective potential near the circular orbit therefore controls the instability timescale through the Lyapunov exponent.

The two phase portraits illustrate photon motion in the radial phase space ( \(r,\dot r\) ) for two situations: the standard Schwarzschild black hole (\(\eta=0\)) and the \(\eta\)-deformed black hole (\(\eta=0.75\)) (see Fig. \ref{Fig8}). The horizontal axis represents the radial distance, while the vertical axis shows the radial velocity of the photon. The streamlines describe how photon trajectories evolve under the gravitational field. Three regions appear in both plots. The red lines correspond to photons that move inward and are captured by the black hole. The teal lines represent photons that are deflected but eventually escape to large distances. Between them lies the critical region (dark blue), where trajectories approach the unstable circular photon orbit. The dashed blue line marks the photon sphere, while the green point represents the saddle point associated with this unstable orbit. The dashed black line indicates the event horizon. Comparing the two cases shows that when \(\eta\) increases, the positions of the photon sphere and the saddle point shift slightly, modifying the structure of the trajectories. This indicates that the deformation parameter affects photon motion and the balance between capture and escape near the black hole.

\section{Observational Implications}\label{Sec:7}
To test the observational viability of the proposed quantum-deformed black hole model, we use the measurements reported by the EHT Collaboration \cite{EventHorizonTelescope:2019dse, EventHorizonTelescope:2019uob, EventHorizonTelescope:2019jan, EventHorizonTelescope:2019ths, EventHorizonTelescope:2019pgp, EventHorizonTelescope:2019ggy, EventHorizonTelescope:2022wkp, EventHorizonTelescope:2022apq, EventHorizonTelescope:2022wok, EventHorizonTelescope:2022exc, EventHorizonTelescope:2022urf, EventHorizonTelescope:2022xqj}. These observations provide a direct probe of the near-horizon region and therefore allow theoretical predictions for the black hole shadow to be confronted with astrophysical data. In our analysis we consider the two supermassive black holes Sgr A$^*$ and M87$^*$, whose geometries can be reasonably approximated as static and spherically symmetric. In the present model, the spacetime is modified by the quantum deformation parameter \( \eta \), which changes the photon-sphere radius and consequently the predicted shadow size. By comparing the theoretical shadow radius with the angular diameters measured by the EHT, we can constrain the allowed values of \( \eta \). This comparison relies on three observational quantities: the angular diameter of the shadow \( \theta \), the distance to the source \( D \), and the black hole mass \( M \), which are well determined for both sources in the EHT data \cite{EventHorizonTelescope:2019dse, EventHorizonTelescope:2019uob, EventHorizonTelescope:2019jan, EventHorizonTelescope:2019ths, EventHorizonTelescope:2019pgp, EventHorizonTelescope:2019ggy,EventHorizonTelescope:2022wkp, EventHorizonTelescope:2022apq, EventHorizonTelescope:2022wok, EventHorizonTelescope:2022exc, EventHorizonTelescope:2022urf, EventHorizonTelescope:2022xqj}.

Recent observations by the EHT Collaboration have provided the first direct images of black hole shadows, opening an important opportunity to test gravitational theories in the strong-field regime. In particular, the observations of the supermassive black holes M87$^*$ and Sgr A$^*$ revealed bright emission rings surrounding a darker central region. This dark area corresponds to the black hole shadow, produced by the capture of photons near the event horizon. The measured angular diameter of the ring is about \(42\pm3\) microarcseconds for M87$^*$ and roughly \(48.7\pm7\) microarcseconds for Sgr A$^*$. These values agree well with theoretical expectations for photon motion in the strong gravitational field close to a black hole. These measurements, combined with independent estimates of the black hole masses and distances, provide the key observational inputs needed to test theoretical models. For M87$^*$, the mass is about \((6.5\pm0.7)\times10^{9} M_\odot\) and the distance is approximately \(16.8\) Mpc. For Sgr A$^*$, the corresponding values are roughly \((4.0-4.3)\times10^{6} M_\odot\) and a distance of about \(8\) kpc. Using these quantities, it is convenient to define a dimensionless shadow diameter per unit mass, \(d_{sh}=D\theta/M\). This quantity removes the dependence on the physical scale of the system and allows different black holes to be compared within the same theoretical framework. From the observational data, the approximate values are \(d_{sh}^{M87^*}\approx(11\pm1.5)M\) and \(d_{sh}^{SgrA^*}\approx(9.5\pm1.4)M\). These estimates provide a direct benchmark for evaluating alternative black hole metrics.

In the present work, the spacetime geometry is described by a quantum-deformed lapse function (\ref{lapsefunction}). The size of the shadow predicted by this model depends on the photon sphere, which is determined from the condition (\ref{PScondition}). Solving the equation (\ref{PScondition}) yields the radius of the unstable photon orbit \(r_{ph}\). The corresponding impact parameter \(b=r_{ph}/\sqrt{f(r_{ph})}\) then defines the theoretical shadow radius that can be compared with observations.

The first numerical analysis (see Fig. \ref{Fig9}, left panel, and Tabs. \ref{tab:shadow_eta} and \ref{tab:EHT_constraints}) studies how the quantity \(b/M\) varies with the deformation parameter \(\eta\). The resulting curve shows how the predicted shadow size changes as the quantum deformation increases. To test the model against observations, horizontal bands corresponding to the \(1\sigma\) and \(2\sigma\) uncertainties from the EHT measurements are added to the plot. The points where the theoretical curve intersects these bands are obtained numerically. Each intersection (\(\eta,b/M\)) represents a value of the deformation parameter for which the predicted shadow size matches one of the observational limits. These points therefore mark the boundaries of the allowed range of \(\eta\). Within this interval, the model remains consistent with the measured shadow size, while outside it the prediction deviates from the observations. Physically, the plot shows that increasing \(\eta\) changes the position of the photon sphere and generally leads to a larger effective shadow radius, reflecting the influence of quantum corrections on the geometry of photon trajectories.

The second numerical analysis (see Fig. \ref{Fig9}, right panel, and Tabs. \ref{tab:shadow_eta} and \ref{tab:EHT_constraints}) extends the study by allowing both the deformation parameter and the black hole mass to vary simultaneously. In this case, the shadow radius becomes a function \(b(\eta,M)\). The observational constraints are imposed through inequalities of the form 
\[R_{obs}-\sigma\le b(\eta,M)\le R_{obs}+\sigma.\]
The resulting diagram in the \((\eta,M)\) plane contains shaded regions representing combinations of parameters that reproduce the observed shadow sizes. Darker areas correspond to the \(1\sigma\) confidence level, while lighter regions indicate the \(2\sigma\) limits. Points located along the boundaries mark the parameter values where the theoretical prediction exactly matches the observational limits. The overlap between the regions derived from Sgr A$^*$ and M87$^*$ identifies the range of parameters for which the quantum-deformed metric is compatible with both astrophysical systems at the same time.

Taken together, these two analyses (see Fig. \ref{Fig9}) provide a clear physical picture. The first left panel shows directly how the deformation parameter modifies the photon capture radius and constrains its value using the observed shadow size. The second right panel reveals how the allowed deformation depends on the black hole mass and highlights the parameter region consistent with both EHT observations. Overall, the results demonstrate that black hole shadow measurements offer a powerful way to test modifications of general relativity. Since the photon sphere and the shadow size depend strongly on the spacetime geometry, the observational results from the EHT Collaboration place meaningful limits on the quantum-deformation parameter and therefore on possible quantum corrections to the black hole metric.

\section{Concluding Remarks}\label{Sec:8}

In this work, we have explored in detail the physical properties of the Kazakov-Solodukhin quantum-deformed black hole spacetime. The model is characterized by a single deformation parameter \( \eta \), which represents quantum corrections to the classical Schwarzschild geometry. When \( \eta \to 0 \), the solution smoothly reduces to the standard Schwarzschild spacetime, fully recovering general relativity. For nonzero values of \( \eta \), however, the geometry changes in an important way, especially in the strong-field region near the event horizon. A major result of this model is the elimination of the classical central singularity. In the Schwarzschild case, curvature invariants diverge at \( r = 0 \). In contrast, the present geometry remains finite near the minimal radius \( r = \eta \). The singular core is therefore replaced by a smooth, regular region. In this way, the parameter \( \eta \) introduces a natural minimal length scale into the spacetime. For values such as \( \eta \sim 0.05M - 0.1M \), the deviations from Schwarzschild become noticeable near the horizon, while the large-distance behavior remains practically unchanged. This makes the model particularly appealing, since it modifies only the strong-gravity regime without disturbing well-tested weak-field predictions. The deformation also influences the horizon structure. As \( \eta \) increases, the event horizon radius shifts outward relative to the classical value \( r_h = 2M \). As a result, the mass-radius relation is modified. Although the effect is small for tiny \( \eta \), it becomes increasingly significant as the deformation grows. This provides a clear geometric picture of how quantum corrections alter black hole structure.

Thermodynamic properties are also affected. The Hawking temperature is proportional to the surface gravity,
\[
T_H = \frac{\kappa}{2\pi} \quad \text{with} \quad \kappa = f'(r_+)/2.
\]
Since the deformation reduces the surface gravity \( \kappa \), the black hole becomes colder as \( \eta \) increases. For example, when \( \eta \sim 0.1M \), the temperature can decrease by a few percent compared with the Schwarzschild case. This leads to a slower evaporation process and a reduced energy emission rate. In practical terms, the quantum-deformed black hole is slightly more stable than its classical counterpart. 

Photon motion provides another important way to understand the impact of the deformation. The effective potential for null geodesics changes, and the photon sphere is determined by
\[
\frac{d}{dr}\!\left(\frac{r^2}{f(r)}\right)=0.
\]
Because this condition depends on \( \eta \), the photon-sphere radius shifts when the deformation is introduced. This modifies the critical impact parameter and affects strong gravitational lensing. Even relatively small values of \( \eta \) can produce noticeable changes in the strong-field region, showing that quantum corrections influence light propagation near compact objects.

The black hole shadow is one of the most important observational signatures of these changes. Although the shadow remains circular due to spherical symmetry, its size depends on the photon sphere. The shadow radius,
\[
R_{sh} = \frac{r_{ph}}{\sqrt{f(r_{ph})}},
\]
varies with \( \eta \). Increasing the deformation parameter alters the shadow size compared to the classical Schwarzschild prediction. Because modern very-long-baseline interferometry observations achieve high angular precision, even small deviations can be tested. This makes the parameter \( \eta \) directly observable in principle. Observational constraints can be expressed by comparing the theoretical shadow radius with the measured value \( R_{obs} \). Consistency with data requires
\[
\left| R_{sh}(\eta) - R_{obs} \right| \leq \Delta R_{obs}.
\]
This inequality provides an upper bound on the deformation parameter,
\[
\eta \leq \eta_{\text{max}},
\]
which translates observational measurements into a physical constraint. In this way, the model becomes testable rather than purely theoretical. In the weak-field limit, the deflection angle receives a correction proportional to \( \eta^2 \). While the leading term reproduces the standard Einstein bending angle, the additional contribution slightly enhances light deflection, particularly for small impact parameters. Importantly, the quadratic dependence ensures that the correction remains very small at large distances, preserving agreement with solar-system tests while allowing measurable effects in strong-lensing environments. Beyond shadow and lensing observations, the model may also be tested through gravitational-wave signals. Since quasinormal mode frequencies depend on the effective potential, any modification of the photon sphere or near-horizon structure can influence the ringdown phase of black hole mergers. Therefore, future gravitational-wave measurements could provide an additional and powerful way to constrain \( \eta \).

Overall, the Kazakov-Solodukhin metric offers a consistent and physically motivated quantum-inspired extension of the Schwarzschild solution. The deformation parameter \( \eta \) removes the central singularity, introduces a minimal length scale, modifies the horizon structure, lowers the Hawking temperature, affects photon motion, and changes both shadow and lensing properties. At the same time, the spacetime reduces smoothly to classical general relativity at large distances, maintaining agreement with well-established gravitational tests. These results show that even a single quantum correction parameter can produce meaningful and potentially observable effects in black hole physics. With improving observational precision in black hole imaging, gravitational lensing, and gravitational-wave astronomy, future data will play a crucial role in testing and constraining such deviations, offering valuable insight into possible quantum aspects of gravity.\\

{
\section*{Acknowledgements} 
We thank the reviewers for their valuable comments and constructive suggestions, which have significantly improved the clarity and quality of this work.}  

\section*{Conflict Of Interest statement } 
No conflict of interest declared by the authors.

\section*{Data Availability Statement:}  
No data were created or analyzed in this study.

\bibliography{references}

@article{EventHorizonTelescope:2019dse,
    author = "Akiyama, Kazunori and others",
    collaboration = "Event Horizon Telescope",
    title = "{First M87 Event Horizon Telescope Results. I. The Shadow of the Supermassive Black Hole}",
    eprint = "1906.11238",
    archivePrefix = "arXiv",
    primaryClass = "astro-ph.GA",
    doi = "10.3847/2041-8213/ab0ec7",
    journal = "Astrophys. J. Lett.",
    volume = "875",
    pages = "L1",
    year = "2019"
}

@article{EventHorizonTelescope:2019uob,
    author = "Akiyama, Kazunori and others",
    collaboration = "Event Horizon Telescope",
    title = "{First M87 Event Horizon Telescope Results. II. Array and Instrumentation}",
    eprint = "1906.11239",
    archivePrefix = "arXiv",
    primaryClass = "astro-ph.IM",
    doi = "10.3847/2041-8213/ab0c96",
    journal = "Astrophys. J. Lett.",
    volume = "875",
    number = "1",
    pages = "L2",
    year = "2019"
}

@article{EventHorizonTelescope:2019jan,
    author = "Akiyama, Kazunori and others",
    collaboration = "Event Horizon Telescope",
    title = "{First M87 Event Horizon Telescope Results. III. Data Processing and Calibration}",
    eprint = "1906.11240",
    archivePrefix = "arXiv",
    primaryClass = "astro-ph.GA",
    doi = "10.3847/2041-8213/ab0c57",
    journal = "Astrophys. J. Lett.",
    volume = "875",
    number = "1",
    pages = "L3",
    year = "2019"
}

@article{EventHorizonTelescope:2019ths,
    author = "Akiyama, Kazunori and others",
    collaboration = "Event Horizon Telescope",
    title = "{First M87 Event Horizon Telescope Results. IV. Imaging the Central Supermassive Black Hole}",
    eprint = "1906.11241",
    archivePrefix = "arXiv",
    primaryClass = "astro-ph.GA",
    doi = "10.3847/2041-8213/ab0e85",
    journal = "Astrophys. J. Lett.",
    volume = "875",
    number = "1",
    pages = "L4",
    year = "2019"
}

@article{EventHorizonTelescope:2019pgp,
    author = "Akiyama, Kazunori and others",
    collaboration = "Event Horizon Telescope",
    title = "{First M87 Event Horizon Telescope Results. V. Physical Origin of the Asymmetric Ring}",
    eprint = "1906.11242",
    archivePrefix = "arXiv",
    primaryClass = "astro-ph.GA",
    doi = "10.3847/2041-8213/ab0f43",
    journal = "Astrophys. J. Lett.",
    volume = "875",
    number = "1",
    pages = "L5",
    year = "2019"
}

@article{EventHorizonTelescope:2019ggy,
    author = "Akiyama, Kazunori and others",
    collaboration = "Event Horizon Telescope",
    title = "{First M87 Event Horizon Telescope Results. VI. The Shadow and Mass of the Central Black Hole}",
    eprint = "1906.11243",
    archivePrefix = "arXiv",
    primaryClass = "astro-ph.GA",
    doi = "10.3847/2041-8213/ab1141",
    journal = "Astrophys. J. Lett.",
    volume = "875",
    number = "1",
    pages = "L6",
    year = "2019"
}

@article{EventHorizonTelescope:2021srq,
    author = "Akiyama, Kazunori and others",
    collaboration = "Event Horizon Telescope",
    title = "{First M87 Event Horizon Telescope Results. VIII. Magnetic Field Structure near The Event Horizon}",
    eprint = "2105.01173",
    archivePrefix = "arXiv",
    primaryClass = "astro-ph.HE",
    reportNumber = "FERMILAB-PUB-21-850-PPD",
    doi = "10.3847/2041-8213/abe4de",
    journal = "Astrophys. J. Lett.",
    volume = "910",
    number = "1",
    pages = "L13",
    year = "2021"
}

@article{EventHorizonTelescope:2022wkp,
    author = "Akiyama, Kazunori and others",
    collaboration = "Event Horizon Telescope",
    title = "{First Sagittarius A* Event Horizon Telescope Results. I. The Shadow of the Supermassive Black Hole in the Center of the Milky Way}",
    eprint = "2311.08680",
    archivePrefix = "arXiv",
    primaryClass = "astro-ph.HE",
    doi = "10.3847/2041-8213/ac6674",
    journal = "Astrophys. J. Lett.",
    volume = "930",
    number = "2",
    pages = "L12",
    year = "2022"
}

@article{EventHorizonTelescope:2022apq,
    author = "Akiyama, Kazunori and others",
    collaboration = "Event Horizon Telescope",
    title = "{First Sagittarius A* Event Horizon Telescope Results. II. EHT and Multiwavelength Observations, Data Processing, and Calibration}",
    eprint = "2311.08679",
    archivePrefix = "arXiv",
    primaryClass = "astro-ph.HE",
    reportNumber = "FERMILAB-PUB-22-418-PPD",
    doi = "10.3847/2041-8213/ac6675",
    journal = "Astrophys. J. Lett.",
    volume = "930",
    number = "2",
    pages = "L13",
    year = "2022"
}

@article{EventHorizonTelescope:2022wok,
    author = "Akiyama, Kazunori and others",
    collaboration = "Event Horizon Telescope",
    title = "{First Sagittarius A* Event Horizon Telescope Results. III. Imaging of the Galactic Center Supermassive Black Hole}",
    eprint = "2311.09479",
    archivePrefix = "arXiv",
    primaryClass = "astro-ph.HE",
    doi = "10.3847/2041-8213/ac6429",
    journal = "Astrophys. J. Lett.",
    volume = "930",
    number = "2",
    pages = "L14",
    year = "2022"
}

@article{EventHorizonTelescope:2022exc,
    author = "Akiyama, Kazunori and others",
    collaboration = "Event Horizon Telescope",
    title = "{First Sagittarius A* Event Horizon Telescope Results. IV. Variability, Morphology, and Black Hole Mass}",
    eprint = "2311.08697",
    archivePrefix = "arXiv",
    primaryClass = "astro-ph.HE",
    reportNumber = "FERMILAB-PUB-22-423-PPD",
    doi = "10.3847/2041-8213/ac6736",
    journal = "Astrophys. J. Lett.",
    volume = "930",
    number = "2",
    pages = "L15",
    year = "2022"
}

@article{EventHorizonTelescope:2022urf,
    author = "Akiyama, Kazunori and others",
    collaboration = "Event Horizon Telescope",
    title = "{First Sagittarius A* Event Horizon Telescope Results. V. Testing Astrophysical Models of the Galactic Center Black Hole}",
    eprint = "2311.09478",
    archivePrefix = "arXiv",
    primaryClass = "astro-ph.HE",
    reportNumber = "FERMILAB-PUB-22-419-PPD",
    doi = "10.3847/2041-8213/ac6672",
    journal = "Astrophys. J. Lett.",
    volume = "930",
    number = "2",
    pages = "L16",
    year = "2022"
}

@article{EventHorizonTelescope:2022xqj,
    author = "Akiyama, Kazunori and others",
    collaboration = "Event Horizon Telescope",
    title = "{First Sagittarius A* Event Horizon Telescope Results. VI. Testing the Black Hole Metric}",
    eprint = "2311.09484",
    archivePrefix = "arXiv",
    primaryClass = "astro-ph.HE",
    reportNumber = "FERMILAB-PUB-22-422-PPD",
    doi = "10.3847/2041-8213/ac6756",
    journal = "Astrophys. J. Lett.",
    volume = "930",
    number = "2",
    pages = "L17",
    year = "2022"
}

@article{Hawking:1970zqf,
    author = "Hawking, S. W. and Penrose, R.",
    title = "{The Singularities of gravitational collapse and cosmology}",
    doi = "10.1098/rspa.1970.0021",
    journal = "Proc. Roy. Soc. Lond. A",
    volume = "314",
    pages = "529--548",
    year = "1970"
}

@article{Duff:1974ud, 
    author = "Duff, M. J.",
    title = "{Quantum corrections to the schwarzschild solution}",
    doi = "10.1103/PhysRevD.9.1837",
    journal = "Phys. Rev. D",
    volume = "9",
    pages = "1837--1839",
    year = "1974"
}

@article{Frolov:1981mz,
    author = "Frolov, Valeri P. and Vilkovisky, G. A.",
    title = "{Spherically Symmetric Collapse in Quantum Gravity}",
    doi = "10.1016/0370-2693(81)90542-6",
    journal = "Phys. Lett. B",
    volume = "106",
    pages = "307--313",
    year = "1981"
}

@article{Stelle:1976gc,
    author = "Stelle, K. S.",
    title = "{Renormalization of Higher Derivative Quantum Gravity}",
    reportNumber = "PRINT-76-1059 (BRANDEIS)",
    doi = "10.1103/PhysRevD.16.953",
    journal = "Phys. Rev. D",
    volume = "16",
    pages = "953--969",
    year = "1977"
}

@article{Stelle:1977ry,
    author = "Stelle, K. S.",
    title = "{Classical Gravity with Higher Derivatives}",
    reportNumber = "Print-77-0417 (BRANDEIS)",
    doi = "10.1007/BF00760427",
    journal = "Gen. Rel. Grav.",
    volume = "9",
    pages = "353--371",
    year = "1978"
}

@article{Biswas:2011ar,
    author = "Biswas, Tirthabir and Gerwick, Erik and Koivisto, Tomi and Mazumdar, Anupam",
    title = "{Towards singularity and ghost free theories of gravity}",
    eprint = "1110.5249",
    archivePrefix = "arXiv",
    primaryClass = "gr-qc",
    doi = "10.1103/PhysRevLett.108.031101",
    journal = "Phys. Rev. Lett.",
    volume = "108",
    pages = "031101",
    year = "2012"
}

@article{Biswas:2013cha,
    author = "Biswas, Tirthabir and Conroy, Aindri{\'u} and Koshelev, Alexey S. and Mazumdar, Anupam",
    title = "{Generalized ghost-free quadratic curvature gravity}",
    eprint = "1308.2319",
    archivePrefix = "arXiv",
    primaryClass = "hep-th",
    doi = "10.1088/0264-9381/31/1/015022",
    journal = "Class. Quant. Grav.",
    volume = "31",
    pages = "015022",
    year = "2014",
    note = "[Erratum: Class.Quant.Grav. 31, 159501 (2014)]"
}

@article{Biswas:2016egy,
    author = "Biswas, Tirthabir and Koshelev, Alexey S. and Mazumdar, Anupam",
    title = "{Consistent higher derivative gravitational theories with stable de Sitter and anti{\textendash}de Sitter backgrounds}",
    eprint = "1606.01250",
    archivePrefix = "arXiv",
    primaryClass = "gr-qc",
    doi = "10.1103/PhysRevD.95.043533",
    journal = "Phys. Rev. D",
    volume = "95",
    number = "4",
    pages = "043533",
    year = "2017"
}

@article{tHooft:1987vrq, 
    author = "'t Hooft, Gerard",
    title = "{Graviton Dominance in Ultrahigh-Energy Scattering}",
    doi = "10.1016/0370-2693(87)90159-6",
    journal = "Phys. Lett. B",
    volume = "198",
    pages = "61--63",
    year = "1987"
}

@article{tHooft:1988oyr,
    author = "'t Hooft, Gerard",
    title = "{On the Factorization of Universal Poles in a Theory of Gravitating Point Particles}",
    doi = "10.1016/0550-3213(88)90659-1",
    journal = "Nucl. Phys. B",
    volume = "304",
    pages = "867--876",
    year = "1988"
}

@article{Battista:2021rlh,
    author = "Battista, Emmanuele and De Falco, Vittorio",
    title = "{First post-Newtonian generation of gravitational waves in Einstein-Cartan theory}",
    eprint = "2109.01384",
    archivePrefix = "arXiv",
    primaryClass = "gr-qc",
    doi = "10.1103/PhysRevD.104.084067",
    journal = "Phys. Rev. D",
    volume = "104",
    number = "8",
    pages = "084067",
    year = "2021"
}

@article{Battista:2023glw,
    author = "Battista, Emmanuele and Steinacker, Harold C.",
    title = "{One-loop effective action of the IKKT model for cosmological backgrounds}",
    eprint = "2310.11126",
    archivePrefix = "arXiv",
    primaryClass = "hep-th",
    doi = "10.1007/JHEP01(2024)125",
    journal = "JHEP",
    volume = "01",
    pages = "125",
    year = "2024"
}

@article{Battista:2023iyu,
    author = "Battista, Emmanuele",
    title = "{Quantum Schwarzschild geometry in effective field theory models of gravity}",
    eprint = "2312.00450",
    archivePrefix = "arXiv",
    primaryClass = "gr-qc",
    doi = "10.1103/PhysRevD.109.026004",
    journal = "Phys. Rev. D",
    volume = "109",
    number = "2",
    pages = "026004",
    year = "2024"
}

@article{Errehymy:2024mlf,
    author = "Errehymy, Abdelghani and Khedif, Youssef and Donmez, Orhan and Daoud, Mohammed and Myrzakulov, Kairat and Bekov, Sabit",
    title = "{Possible wormholes in $f(R)$ gravity sourced by solitonic quantum wave and cold dark matter halos and their repulsive gravity effect}",
    eprint = "2408.07667",
    archivePrefix = "arXiv",
    primaryClass = "gr-qc",
    doi = "10.1140/epjc/s10052-024-13224-4",
    journal = "Eur. Phys. J. C",
    volume = "84",
    number = "9",
    pages = "908",
    year = "2024"
}

@article{Battista:2024gud,
    author = "Battista, Emmanuele and Capozziello, Salvatore and Errehymy, Abdelghani",
    title = "{Generalized uncertainty principle corrections in Rastall{\textendash}Rainbow Casimir wormholes}",
    eprint = "2409.09750",
    archivePrefix = "arXiv",
    primaryClass = "gr-qc",
    doi = "10.1140/epjc/s10052-024-13656-y",
    journal = "Eur. Phys. J. C",
    volume = "84",
    number = "12",
    pages = "1314",
    year = "2024"
}

@article{Errehymy:2025llh,
    author = "Errehymy, A. and Turimov, B. and Syzdykova, A. and Myrzakulov, K. and Alessa, N. and Abdel-Aty, A. -H.",
    title = "{Dymnikova-Schwinger GUP-corrected wormholes in f(R,Lm,T) gravity}",
    doi = "10.1016/j.nuclphysb.2025.117116",
    journal = "Nucl. Phys. B",
    volume = "1019",
    pages = "117116",
    year = "2025"
}

@article{Wang:2025fmz,
    author = "Wang, Zi-Liang and Battista, Emmanuele",
    title = "{Dynamical features and shadows of quantum Schwarzschild black hole in effective field theories of gravity}",
    eprint = "2501.14516",
    archivePrefix = "arXiv",
    primaryClass = "gr-qc",
    doi = "10.1140/epjc/s10052-025-13833-7",
    journal = "Eur. Phys. J. C",
    volume = "85",
    number = "3",
    pages = "304",
    year = "2025"
}

@article{Errehymy:2025djk,
    author = "Errehymy, A. and Khedif, Y. and Daoud, M. and Myrzakulov, K. and Turimov, B. and Myrzakul, T.",
    title = "{Quantum corrections to Dymnikova-Schwinger black holes in Einstein-Gauss-Bonnet gravity}",
    eprint = "2509.17630",
    archivePrefix = "arXiv",
    primaryClass = "gr-qc",
    doi = "10.1016/j.physletb.2025.139915",
    journal = "Phys. Lett. B",
    volume = "870",
    pages = "139915",
    year = "2025"
}

@article{Kazakov:1993ha,
    author = "Kazakov, D. I. and Solodukhin, S. N.",
    title = "{On Quantum deformation of the Schwarzschild solution}",
    eprint = "hep-th/9310150",
    archivePrefix = "arXiv",
    reportNumber = "JINR-E2-93-371",
    doi = "10.1016/S0550-3213(94)80045-6",
    journal = "Nucl. Phys. B",
    volume = "429",
    pages = "153--176",
    year = "1994"
}

@article{Konoplya:2019xmn,
    author = "Konoplya, R. A.",
    title = "{Quantum corrected black holes: quasinormal modes, scattering, shadows}",
    eprint = "1912.10582",
    archivePrefix = "arXiv",
    primaryClass = "gr-qc",
    doi = "10.1016/j.physletb.2020.135363",
    journal = "Phys. Lett. B",
    volume = "804",
    pages = "135363",
    year = "2020"
}

@article{Lu:2021htd, 
    author = "Lu, Xu and Xie, Yi",
    title = "{Gravitational lensing by a quantum deformed Schwarzschild black hole}",
    doi = "10.1140/epjc/s10052-021-09440-x",
    journal = "Eur. Phys. J. C",
    volume = "81",
    number = "7",
    pages = "627",
    year = "2021"
}

@article{Bezerra:2019qkx, 
    author = "Bezerra, V. B. and Lobo, I. P. and Morais Gra{\c{c}}a, J. P. and Santos, Luis C. N.",
    title = "{Effects of quantum corrections on the criticality and efficiency of black holes surrounded by a perfect fluid}",
    eprint = "1908.08140",
    archivePrefix = "arXiv",
    primaryClass = "gr-qc",
    doi = "10.1140/epjc/s10052-019-7482-0",
    journal = "Eur. Phys. J. C",
    volume = "79",
    number = "11",
    pages = "949",
    year = "2019"
}

@article{Berry:2021hos, 
    author = "Berry, Thomas and Simpson, Alex and Visser, Matt",
    title = "{General class of ''quantum deformed'' regular black holes}",
    eprint = "2102.02471",
    archivePrefix = "arXiv",
    primaryClass = "gr-qc",
    doi = "10.3390/universe7060165",
    journal = "Universe",
    volume = "7",
    number = "6",
    pages = "165",
    year = "2021"
}

@article{Gao:2021ybp, 
    author = "Gao, Bo and Deng, Xue-Mei",
    title = "{Dynamics of charged test particles around quantum-corrected Schwarzschild black holes}",
    doi = "10.1140/epjc/s10052-021-09782-6",
    journal = "Eur. Phys. J. C",
    volume = "81",
    number = "11",
    pages = "983",
    year = "2021"
}

@article{Javed:2021ymu, 
    author = {Javed, Wajiha and Hussain, Iqra and {\"O}vg{\"u}n, Ali},
    title = "{Weak deflection angle of Kazakov{\textendash}Solodukhin black hole in plasma medium using Gauss{\textendash}Bonnet theorem and its greybody bonding}",
    eprint = "2201.09879",
    archivePrefix = "arXiv",
    primaryClass = "gr-qc",
    doi = "10.1140/epjp/s13360-022-02374-7",
    journal = "Eur. Phys. J. Plus",
    volume = "137",
    number = "1",
    pages = "148",
    year = "2022"
}

@article{Peng:2020wun,
    author = "Peng, Jun and Guo, Minyong and Feng, Xing-Hui",
    title = "{Influence of quantum correction on black hole shadows, photon rings, and lensing rings}",
    eprint = "2008.00657",
    archivePrefix = "arXiv",
    primaryClass = "gr-qc",
    doi = "10.1088/1674-1137/ac06bb",
    journal = "Chin. Phys. C",
    volume = "45",
    number = "8",
    pages = "085103",
    year = "2021"
}

@article{Page:1976df,
    author = "Page, Don N.",
    title = "{Particle Emission Rates from a Black Hole: Massless Particles from an Uncharged, Nonrotating Hole}",
    doi = "10.1103/PhysRevD.13.198",
    journal = "Phys. Rev. D",
    volume = "13",
    pages = "198--206",
    year = "1976"
}

@article{Gibbons:2008rj,
    author = "Gibbons, G. W. and Werner, M. C.",
    title = "{Applications of the Gauss-Bonnet theorem to gravitational lensing}",
    eprint = "0807.0854",
    archivePrefix = "arXiv",
    primaryClass = "gr-qc",
    doi = "10.1088/0264-9381/25/23/235009",
    journal = "Class. Quant. Grav.",
    volume = "25",
    pages = "235009",
    year = "2008"
}

@article{Huo:2023,
    author = "Huo, M. and Fan, Z. and Qi, J. and Qi, N. and Zhu, D.",
    title = "{Fast Analysis of Multi-Asteroid Exploration Mission Using Multiple Electric Sails}",
    doi = "10.2514/1.G006972",
    journal = "J. Guid. Control Dyn.",
    volume = "46",
    number = "5",
    pages = "1015--1022",
    year = "2023"
}

@article{Yu:2025,
    author = "Yu, W. and Wu, Q. and Chen, X. and Huo, J. and Li, J. and Yang, J. and Zhang, A. et al.",
    title = "{Experimental First-Photon Visualization of Quantum Erasure With Hybrid Entanglement}",
    doi = "10.1002/lpor.202501816",
    journal = "Laser Photonics Rev.",
    pages = "e1816",
    year = "2025"
}

@article{Fang:2025,
    author = "Fang, Y. and Cai, R.",
    title = "{Probing the Merger Rates of Supermassive Black Holes and Galaxies with Gravitational Waves}",
    doi = "10.1093/mnras/staf1278",
    journal = "Mon. Not. Roy. Astron. Soc.",
    volume = "542",
    number = "2",
    pages = "1172--1187",
    year = "2025"
}

@article{Qiao:2016,
    author = "Qiao, Z. and Deng, X. and Cubuk, E. D. and Chen, H. and Zhu, W. and Kaxiras, E. and Qi, S. et al.",
    title = "{High-Temperature Quantum Anomalous Hall Effect in $n$-$p$ Codoped Topological Insulators}",
    doi = "10.1103/PhysRevLett.117.056804",
    journal = "Phys. Rev. Lett.",
    volume = "117",
    number = "5",
    pages = "056804",
    year = "2016"
}

@article{Xiang:2026,
    author = "Xiang, Y. and Ye, Y. and Feng, P. and Li, H. and Pang, X. and Lan, X. and Zhang, N. et al.",
    title = "{A Novel Pipeline for the Identification of New Gamma-Ray Blazars from the 4FGL--Xiang Catalog Based on Multiwavelength Flux Distributions}",
    doi = "10.1093/mnras/stag150",
    journal = "Mon. Not. Roy. Astron. Soc.",
    volume = "546",
    number = "4",
    pages = "stag150",
    year = "2026"
}

@article{Zhang:2026,
    author = "Zhang, C. and Lou, J. and Zhang, J. and Wang, Z. and Ji, C. and Yuan, H. and Chang, C. et al.",
    title = "{Space-Time Wavefront Synchronized Terahertz Metasurface}",
    doi = "10.1002/adma.202520890",
    journal = "Adv. Mater.",
    volume = "38",
    number = "16",
    pages = "e20890",
    year = "2026"
}

@article{Errehymy:2026hbj,
    author = "Errehymy, A. and Khedif, Y. and Daoud, M. and Turimov, B. and Usanov, S. and Turaev, F. and Yasakov, Z.",
    title = "{Constraints on ultralight bosonic dark matter soliton halos around Schwarzschild black holes from high-frequency quasi-periodic oscillations in X-ray binaries}",
    doi = "10.1016/j.physletb.2026.140552",
    month = "5",
    year = "2026"
}

@article{Errehymy:2026ftm,
    author = "Errehymy, A. and Khedif, Y. and Daoud, M. and Myrzakulov, Y. and Donmez, O. and Turimov, B.",
    title = "{Event horizon telescope observational constraints on Dymnikova-type non-singular black holes in higher dimensions}",
    eprint = "2601.06711",
    archivePrefix = "arXiv",
    primaryClass = "gr-qc",
    doi = "10.1016/j.physletb.2026.140168",
    journal = "Phys. Lett. B",
    volume = "873",
    pages = "140168",
    year = "2026"
}

@article{Errehymy:2025ada,
    author = "Errehymy, A. and Hansraj, S. and Hansraj, C.",
    title = "{Black-hole Shadows and Null Geodesics in Hamaus{\textendash}Sutter{\textendash}Wandelt Void Spacetimes with a Quintessential Field: Observational Signatures from EHT Data of M87$^{∗}$ and Sgr A$^{∗}$}",
    doi = "10.3847/1538-4357/ae197f",
    journal = "Astrophys. J.",
    volume = "995",
    number = "2",
    pages = "148",
    year = "2025"
}

@article{Errehymy:2025zuj,
    author = "Errehymy, A. and Mustafa, G. and Rasheed, A. S. and Boshkayev, K. and Maurya, S. K. and Alessa, N. and Abdel-Aty, A. H.",
    title = "{Circular orbits and observational signatures in regular black hole spacetimes}",
    doi = "10.1142/S0219887826500180",
    journal = "Int. J. Geom. Meth. Mod. Phys.",
    volume = "23",
    number = "02",
    pages = "2650018",
    year = "2026"
}

@article{Errehymy:2025nuv,
    author = "Errehymy, A. and Hansraj, S. and Myrzakulov, Y.",
    title = "{Einstein-Gauss-Bonnet black holes seeded by cosmic voids}",
    doi = "10.1016/j.physletb.2025.139962",
    journal = "Phys. Lett. B",
    volume = "870",
    pages = "139962",
    year = "2025"
}

@article{Errehymy:2025pfr,
    author = "Errehymy, A.",
    title = "{Null geodesics and shadows of Dekel-Zhao-type dark matter black holes with a quintessential field: Constraints from EHT observations of M87* and Sgr A*}",
    doi = "10.1016/j.physletb.2025.139945",
    journal = "Phys. Lett. B",
    volume = "870",
    pages = "139945",
    year = "2025"
}

@article{Errehymy:2023xpc,
    author = "Errehymy, Abdelghani and Maurya, S. K. and Mustafa, G. and Hansraj, Sudan and Alrebdi, H. I. and Abdel-Aty, Abdel-Haleem",
    title = "{Black Hole Solutions with Dark Matter Halos in the Four-Dimensional Einstein-Gauss-Bonnet Gravity}",
    doi = "10.1002/prop.202300052",
    journal = "Fortsch. Phys.",
    volume = "71",
    number = "10-11",
    pages = "2300052",
    year = "2023"
}

\end{document}